\documentstyle[12pt,aaspp4]{article}

\def\gtsim{\ {\raise-0.5ex\hbox{$\buildrel>\over\sim$}}\ }
\def\ltsim{\ {\raise-0.5ex\hbox{$\buildrel<\over\sim$}}\ }
\slugcomment{The Astronomical Journal, accepted for publication}

\begin{document}
\singlespace

\title{
A dwarf irregular galaxy at the edge of the Local Group:
Stellar populations and distance of IC 5152\footnote{
Based on observations collected
at the La Silla Observatory, operated by the Eorpean Southern Observatory,
and on archival data of the NASA/ESA Hubble
Space Telescope, which is operated by AURA, Inc., under NASA contract
NAS5--26555.}}

\author{Albert A. Zijlstra $^{1}$, Dante Minniti $^{2,3}$}

\altaffiltext{1}{Department of Physics, UMIST, P.O.Box 88, 
 Manchester M60 1QD, UK\\
E-mail: aaz@iapetus.phy.umist.ac.uk}

\altaffiltext{2}{Lawrence Livermore National Laboratory, MS L-413, 
P.O. Box 808,Livermore, CA 94550\\
E-mail:  dminniti@llnl.gov}

\altaffiltext{3}{Departmento de Astronomia, P. Universidad Catolica,
Casilla 104, Santiago 22, Chile}

\begin{abstract}
We have obtained V and I-band photometry for about 700 stars in the
field of the dwarf irregular galaxy IC 5152, previously considered a
possible member of the Local Group.  Deep VI color--magnitude diagrams
are presented of the main body of this galaxy and of a nearby field.
We infer a distance by comparison with the VI color--magnitude
diagrams of similar galaxies and with theoretical isochrones.  The
distance modulus to IC~5152 is found to be $m-M_o = 26.15 \pm 0.2$,
adopting $E(V-I) = 0$.  This distance of $D=1.7$ Mpc implies that
IC~5152 is not a certain member of the Local Group.  We also construct
a deep optical luminosity function. By comparison with theoretical
isochrones, we find a metallicity $Z\approx 0.002$, which is lower
than previous estimates.  Using this metallicity and distance, IC 5152
now fits the metallicity--luminosity relation for dwarf galaxies. The
youngest stars in the field studied have an age of $\sim 10^{7}$ Gyr,
and there is a substantial population of stars with $\log~t>7.8-8$
yr. The central region of IC~5152 is an active site of star formation,
as found by comparing our optical photometry with Hubble Space
Telescope ultraviolet images. Several candidate globular clusters are
found, as well as a candidate for a potential nucleus.
\end{abstract}


\section{Introduction}

Dwarf galaxies are the most numerous galaxies in the Local Group and
nearby Universe. Their evolution shows evidence of being affected by
their environment, with gas-poor dwarf spheroidals mainly occurring
near larger galaxies, and gas-rich dwarf irregulars being more common
in relative isolation (e.g. van den Bergh 1994a).  In return, dwarf
galaxies may themselves have contributed to the formation of spiral
galaxies such as the Milky Way, because gravitational clustering on
progressively larger scales can have played an important role in the
formation of large galaxies (White \& Rees 1978). In this context, it
is important to study the isolated dwarfs, that live in low density
environments, distant enough from other major galaxies to have
remained unperturbed during a Hubble time.  

The Sdm IV--V galaxy IC 5152 is one such system, previously considered
to be a member of the Local Group.  Sandage (1986) quotes a distance
modulus of $m-M_0=24.9$ from the brightest supergiants, equivalent to
$D=1.58$ Mpc.  Recently, however, van den Bergh (1994a) points out
that the distances measured, ranging from 1.5 Mpc to 3.0 Mpc
(Bottinelli et al. 1984), are too uncertain to secure membership.  A
summary of parameters for this galaxy is given in Table 1.  Despite
the fact that IC~5152 is listed as one of the galaxies most easily
resolved into stars (Sandage \& Bedke 1985), there is no previous
color--magnitude diagram of this galaxy, and its past star formation
history is unknown (Mateo 1998, Grebel 1998). A very bright star in
front of this galaxy (HD209142, with $V=7.9$) makes it difficult to
obtain deep exposures (see Figure 1).  Bedding et al.  (1997) took
advantage of the bright star in the field as a reference for using an
adaptive optics system, in order to resolve the galaxy into stars.
They could only detect a few stars at very faint levels, and concluded
that this galaxy must be more distant than previously thought.

In this paper we discuss the stellar populations of the isolated dwarf
irregular galaxy IC~5152, based on VI photometry (down to I$ = 23.0$)
covering a large area to the SE of the main body of this galaxy.
{}From the VI color--magnitude diagram, and a deep luminosity
function, we measure a distance of 1.7 Mpc to IC~5152, which places it
at the edge of the Local Group.

The observations of IC 5152, data reductions, and photometry are
described in Section 2. The resulting color-magnitude diagrams and
luminosity fuctions are given in Section 3, along with a discussion
about reddening.  Fundamental parameters of IC 5152 (metallicity, age
and distance) are determined in Section 4.  Other population tracers
(star clusters, carbon stars, HII regions, planetary nebulae, Cepheid
variables, CO and HI observations) are reviewed in Section 5.  The
possible association of IC~5152 with either the Local Group or the
Sculptor Group is discussed in Section 6.  Finally, the conclusions of
this work are summarized in Section 7.

\section {The Data}

\subsection {Observations and Reductions}

The observations of IC 5152 were obtained during the night of
September 5, 1995, as part of a long term monitoring program of the
variable stars in several galaxies (Zijlstra, Minniti \& Brewer 1997).  We
used the red arm (RILD mode) of the ESO Multi-Mode Instrument (EMMI)
at the New Technology telescope (NTT) in La Silla, operated in active
remote control (Zijlstra et al. 1997) from ESO headquarters in
Garching, Germany.  The weather was photometric.
We used the 2048$\times$2048 Tek
CCD, with a scale of $0.268\arcsec$ pix$^{-1}$.

The observations of this galaxy consist of pairs of frames with
exposure times of 900 sec in I (airmass 1.1), and 1800 sec in V
(airmass 1.1).  A total of 5 standards of Landolt (1992) were observed
during the night in order to calibrate our photometry.

The reductions of the CCD frames were carried out following standard
procedures, using the package CCDRED within the IRAF environment. 
The pairs of images in each filter were combined, the final FWHM
of stellar images being 0.8 arcsec in V and I.

Because of the superposition of a very bright foreground star
(HD209142 = SAO247284, with $V=7.9$, $B-V=0.1$; see Figure 1), this
galaxy has not been observed systematically before. We pointed the
telescope away from the bright star, covering the SE corner of
IC5152.  Nonetheless, additional reflections of this star in the telescope
and instrument optics still cause a residual gradient in the background, 
which ultimately presents a problem when flat
fielding the data.  Therefore, we adopt a conservative systematic error 
of 0.1 mag in all the photometry derived here. Also, even though the 
field covers $8.5\times 9$ arcmin, we use only a region of 
$9\times 3$ arcmin, covering about one quarter of this
galaxy plus a large enough neighboring field for comparison (Figure 2).

\subsection{The Photometry}

The photometric transformations to the standard system were done
following the procedure described by Minniti \& Zijlstra (1997).
These transformations reproduce the magnitudes of the standard stars
with an $rms$ of 0.02 in V and 0.03 in I.

The photometric measurements were performed within the IRAF
environment, using DAOPHOT II which is an improved version of the
original DAOPHOT package developed by Stetson (1987).  In particular,
DAOPHOT II can accommodate a spatially varying point-spread function
across the field.  All stars in the field of IC 5152 with more than
4$\sigma$ above the background in the V and I frames were located and
their magnitudes measured by fitting an empirical point-spread
function.  The position of these objects is plotted schematically in
Figure 3.  Figure 4 shows the DAOPHOT errors as function of $V$ and
$I$-band magnitudes, excluding the flat fielding uncertainty.
The resulting limiting magnitudes at 4$\sigma$
are V$= 24$, and I$= 23$.

The completeness in the photometry varies across the field.  As
expected, the completeness is worse in the more crowded disk region of
IC 5152 ($C \sim 75\%$) than in the outer regions ($C \sim 90\%$).
However, the present photometry is sufficiently deep that none of our
results will depend on the completeness of the sample at the faintest
magnitudes. Most of the stars detected in the I-frames down to $I =
23.0$ are concentrated in the disk of the galaxy.  IC~5152 is
reasonably face on, with an inclination $i=55^{\circ}$ (Hutchmeier \&
Richter 1986).

The photometry of the IC~5152 disk field is listed in Table 2.
Columns 1 and 2 give the star positions in pixels.  With the scale of
$0.268$ arcsec/pix, $1$ arcmin is equivalent to $0.5$ kpc at the
distance of $D=1.7$ Mpc.  Columns 3 to 6 give the V and I magnitudes,
with their respective errors. Finally, column 6 lists the $V-I$
colors.

\section{The color--magnitude diagrams and luminosity function}

\subsection{The Color--Magnitude Diagrams}

Deep optical color--magnitude diagrams are the main tools to study the
stellar content of resolved galaxies (e.g. Greggio et al. 1993, Lee
1993, Tosi 1994, Tolstoy 1995, Tolstoy \& Saha 1996, Marconi et
al. 1994, Gallart et al. 1994, 1996, Dohm-Palmer et al. 1997). The
lack of a deep color-magnitude diagram for IC5152 has caused the star
formation history of this galaxy to remain unknown (see Mateo 1998,
Grebel 1998). Some photographic photometry exists in the literature:
Sandage (1986) quotes distance results from photometric measurements
of the brightest giants in this galaxy.

Figure 5 shows  the first color-magnitude diagrams based on CCD photometry
of the dwarf galaxy IC~5152. It shows the $I ~vs ~V-I$, and $V
~vs ~V-I$ color-magnitude diagrams for a total of 534 stars in the SE
portion of the IC 5152 disk, covering $6.7$ arcmin$^2$. These are all
the stars with centroids matched in V and I frames to better than 2.0
pixels ($0.6\arcsec$), and satisfying stringent criteria of
photometric quality ($\sigma \leq 0.5$, $\chi \leq 2$, and Sharpness
$\geq -1$). The sharpness criterion eliminates galaxies as well as
possible star clusters.  Contamination from foreground stars
has not been subtracted from this diagram; this contamination is
discussed in the next Section.

In spite of the relatively small number of stars, a number of features
can be identified in Figure 5. A tail of red giant stars is seen
reaching $I=18.5$ (we will argue below that this traces red
supergiants rather than the RGB). A blue main sesquence with mean color
$V-I = 0.0$ also reaches $I=18.5$.  A population of blue loop stars
may also be seen, running in a sequence about 0.3 mag redder, parallel
to the main sequence.

\subsection{Reddening}

Because of the high Galactic latitude, $b=-50.19^{\circ}$, the
foreground reddening towards IC~5152 is likely to be small.  Based on
the HI column density, Burstein \& Heiles (1984) estimate the
foreground reddening as $E(B-V)=0.00$.

The bright foreground star HD209142 gives an additional constraint on
the reddening. Photometry from the ESA Tycho catalog (Hog et al. 1998)
gives $V=7.78$ and $B=8.01$ for this A8V star.  This is precisely the
color expected for its type and indicates $E(B-V)<0.02$. The star is
at approximately 120 pc distance which at this Galactic latitude is
close to the edge of the absorbing gas layer in the Galactic disk.
This result is consistent with the negligible reddening inferred by Burstein
\& Heiles (1984).

We can also estimate the foreground reddening by using the sharp blue
cutoff in the stellar distribution of the outer IC 5152 fields.  This
cutoff is interpreted as the locus of the main sequence, which is
nearly vertical in the theoretical color--magnitude diagrams, having
an intrinsic color of $(V-I)_0 = -0.20$ (Bertelli et al. 1994).
{}From the observed color we deduce $E(V-I) < 0.03$, consistent with
the $op.~cit.$ value.  $E(B-V)<0.02$ is equivalent to $E(V-I) < 0.03$
and $A_I < 0.03$, using the reddening ratios $E(V-I) = 1.60 ~E(B-V)$,
$A_I = 1.49~ E(B-V)$ of Rieke \& Lebofski (1985).  We will adopt
$E(V-I) = 0$, and $A_I = 0$.

The main sequence appears broadened more than expected from the
photometric errors, from which we infer that there is substantial
internal differential absorption in IC~5152.  Webster et al. (1983)
derive $c_{H\beta}=0.23$ for one HII region, corresponding to $E(V-I)
=0.25$, and indicating the presence of internal extinction near this
HII region.  Although internal extinction is (on average) less
important for red giants than for luminous main sequence stars, as the
latter occur closer to the dusty star forming regions, the effect also
broadens the giant branches at fainter magnitudes.

\subsection{Contamination from Foreground Stars and Background Galaxies}

The foreground stars from the Milky Way halo and disk should appear as
a plume of stars with $V-I = 0.7$, with total range covering about
$0.4 \leq V-I \leq 1.5$.  We divide the observed $9\times 3$ arcmin
field in two zones, one containing the IC~5152 disk covering $6.7$
arcmin$^2$, and the rest containing mostly foreground stars of our own
galaxy and background galaxies, covering $16.5$ arcmin$^2$.  The
photometry of $N=144$ objects in the neighboring field is listed in
Table 3, and their color-magnitude diagram is shown in Figure 6.  The
observed density of red stars ($V-I > 1.0$) in this region is $1$
stars/arcmin$^2$ with $I\leq 20$, and $2$ stars/arcmin$^2$ with $I
\leq 21$.  Thus, in the location of the supergiant branch of IC~5152
of Figure 5, we expect no more than about 14 objects that may be
non-members of this galaxy.

The few stars of Figure 6 located in the region corresponding to the
IC~5152 supergiant branch may still be members of this galaxy if the
IC~5152 disk is more extended than the region adopted above.
{}From the Galactic model of Ratnatunga \& Bahcall (1985), we expect
0.7 foreground halo stars/arcmin$^2$ with $V \leq 21$, and $0 \leq V-I
\leq 2$ in this high-latitude field ($l=343.91^{\circ},
~b=-50.19^{\circ}$).

Since the galaxy/star ratio increases rapidly for faint magnitudes,
the background galaxy contamination also has to be taken into account.
Most of the brighter galaxies are resolved and are discarded by our
sharpness criterion; only a few fainter and very compact ones may
cause confusion. {}From Tyson (1988) we expect a galaxy density of
$10^4\,$gal$\,$degree$^{-2}$mag$^{-1}$ with $I<21.5$. This would give
$\sim 75$ galaxies in our $3\times9$ arcmin$^2$ field, and only $\sim
22$ galaxies in the $6.7$ arcmin$^2$ of IC~5152 corresponding to the
color-magnitude diagram shown in Figure 5.  However, the majority of
these galaxies are resolved, and were eliminated on that basis.  In
summary, the field contamination would not alter the appearance of
Figure 5, which truly represents the stellar content of the IC~5152
disk.

\subsection{The Luminosity Function}

An I-band luminosity function can be constructed by counting all the
stars present in the I-frames, regardless of them matching the
V-images.  The normalized counts of $N=770$ stars measured in the
I-band are plotted in Figure 7. Two distinct breaks can be seen: the
brightest at $I\approx 18.5\pm 0.1$, and another break at $I = 20.4$.
The luminosity function corresponding to the foreground stars and
unresolved background galaxies is plotted in Figure 8 for comparison.

A bolometric luminosity function can be constructed for the red stars
with $V-I>0.8$, using the following relation:
$$M_{bol}=I_0+0.30+0.38(V-I)_0-0.14(V-I)_0^2-(m-M)_0$$ (Bessell \&
Wood 1984). The resulting bolometric luminosity function of the giant
branches is shown in Figure 9. The two breaks are seen very clearly,
at $m_{bol} = 16.5$ and $m_{bol} = 19$.

\subsection{ Interpretation}

The sequence of red stars can contain red Supergiants, AGB stars and
RGB stars; the two breaks indicate at least two of these are
present. Because the Supergiant and/or the AGB break may not be
obvious in a small system, a consistent interpretation could be
possible at several distances.  For a sparse color--magnitude diagram,
further information is needed to remove the distance ambiguity
(e.g. Tolstoy et al. 1998).

The tip of the RGB is an excellent distance indicator\footnote{A sparsely populated RGB can hide the actual tip of the RGB, giving
the appearance of a break at a too faint magnitude. In this case the RGB
loses its power as distance indicator, and the derived distance 
becomes an upper limit.}
 (e.g. Lee 1993),
having $M_I=-4.0$ for old and metal-poor populations. IC~5152 is known
to have a low metallicity from its HII regions: $12+\log(O/H)=7.71$
(Hidalgo-G\'amez \& Olofsson 1998; see below).  If the break at
$I\approx 18.5\pm 0.1$ were the RGB, the distance would be $\sim
300\,$kpc, whilst the second break at $I = 20.4$ would yield
$760\,$kpc. Both can be considered unlikely, for the following
reasons:

\noindent  1. The failure of adaptive optics observations to resolve
stars in this galaxy gave a lower limit to the distance of 1.5 Mpc
(Bedding et al. 1997). The Cepheids discovered by Caldwell et
al. (1988) also rule out a small distance. \\
2. The presence of HII regions and UV bright
stars (see below) indicate the presence of a young
population. Therefore the stellar population is expected to include
young red supergiants. \\
 3. At these distances IC~5152 would be the smallest dIrr known ($M_B =
-10.7$ and $-12.7$, respectively) and would be located away from the
mean relation defined by its peers in the $L$ versus [Fe/H] plane.

We conclude that a consistent interpretation within the available
data is reached if the two breaks are identified with the tip of the
Red Supergiant sequence and the tip of the AGB.  At a distance of
$m-M_0=26.15$ (see below), the tip of the RGB would be found at
$I=22.2$ for $E_{V-I}=0$, hidden by the incompleteness in the
photometry.

\section{Fundamental parameters of IC 5152}

\subsection{The Distance and Total Luminosity of IC~5152}

In the absence of a detected tip of the RBG, other, less accurate
distance indicators have to be used.  We obtain the distance to
IC~5152 by direct comparison with the color--magnitude diagrams of
similar galaxies: the LMC (Shapley Constellation III, Reid et
al. 1987), the SMC (Reid \& Mould 1990), WLM (Minniti \&
Zijlstra 1997), and NGC~3109 (Minniti et al. 1998). These galaxies
bracket the likely metallicity of IC~5152, as well as its total
luminosity.  The comparison with the last two galaxies is shown in
Figure 10.

We find a difference $\Delta \mu$ of
distance modulus of $7.6$, $1.4$, and $0.5$ mag with the LMC, WLM, and
NGC~3109, respectively. For $A_I = 0$, the distance modulus to IC 5152
is $m-M_0 = 26.15 \pm 0.2$, which translates into $D = 1.7$ Mpc. The
error bar in the distance modulus includes a possible systematic
uncertainty of 0.1 mag in the photometric zeropoint.  The total
luminosity for IC~5152 has been found to be $M_B=-14.5$ for
$m-M_0=25.9$ (e.g. Huchtmeier \& Richter 1986). With the present, more
accurate distance determination, this should be revised to
$M_B=-14.8$.

With this distance, the brightest red supergiants that are likely to
be members of IC~5152 reach $M_{bol}=-9.5$.  The termination of the
AGB corresponds to $M_{bol}=-7.5$ to $-7.0$, in approximate agreement
with the theoretical limit ($M_{bol}=-7.2$) for the most massive AGB
stars. The termination of AGB stars near this limit is also seen in
the LMC (e.g. Zijlstra et al. 1996). The observed range for the most
luminous red supergiants is also in accordance with expectations
(e.g. Chiosi \&\ Maeder 1986) implying progenitor masses up to 50
$M_\odot$.

At the low metallicity, this AGB may be dominated by carbon stars.
The relation used for the bolometric magnitude is slightly
inappropriate for carbon stars (Brewer et al. 1995), but the carbon
star contribution would only be important for the faintest magnitudes,
where we are incomplete.  At low metallicities, the brightest AGB
stars may in fact again be oxygen rich due to Hot Bottom Burning,
whereas lower on the AGB carbon stars dominate (e.g. van Loon et
al. 1998). The peak AGB luminosity indicated in IC~5152 corresponds to
a population of AGB stars with age $\sim 10^8$ yr.

\subsection{Metal Abundances and Ages}

The metallicity of old and metal-poor populations can be derived from
the V$-$I color of their RGB (e.g. Da Costa \& Armandroff 1990). A
similar technique can be applied to the color of the supergiant stars,
since the theoretical isochrones predict that these stars will also
get bluer as the metallicity decreases (Bertelli et al. 1994, Bressan
et al. 1994).  Unfortunately, metallicity and age are degenerate,
producing similar effects in the optical color-magnitude diagrams.  In
order to avoid systematics, we then compare the color of the stars in
the supergiant branch of IC~5152 with those of similar galaxies. As
before, we chose the LMC, SMC, WLM and NGC~3109, which bracket the
metallicity of IC~5152.  The IC~5152 stars are about 0.05 mag redder
than the corresponding stars in WLM and NGC~3109 (Fig. 10), indicating
that they are more metal-rich in the mean.

The ages of old and intermediate age stellar populations can be
derived using the absolute magnitudes of their brightest red stars.
If the photometry reaches beyond the red giant branch, the oldest
populations can be studied (e.g. Lee 1993, Minniti \& Zijlstra 1996,
1997).  Since this is not the case here, we can only study the
population of intermediate-age stars in the IC~5152 disk.  Figure 7
shows the $I ~vs ~V-I$ color-magnitude diagrams for the SE portion of
the IC 5152 disk, along with the isochrones of Bertelli et al. (1994)
corresponding to two different metallicities: $Z/Z_{\odot}=0.05$ and
$0.2$. In order to plot the isochrones, we have assumed $m-M=26.15$
and $E_{V-I}=0$.  

The color of the main sequence agrees with the isochrones, indicating
that the reddening is nearly zero. The brightest stars are consistent
with isochrones of $\log~t=7.0$ yr. The colors and magnitudes of the
well populated red supergiant branch overlaps the isochrones of
$\log~t=7.8-8.0$ yr for $Z=0.001$, being slightly bluer than the
corresponding $Z=0.004$ isochrones.  Stars with intermediate colors
(yellow stars with $0.2<V-I<1.0$) should be stars doing blue loops
according to the isochrones. These yellow stars appear in large
numbers at $I>21.0$. Above $I=21.0$ there is a clear separation
between the main sequence and the supergiant branches. This is
evidence of a large population of stars with ages $\log t>7.8$ yr,
consistent with the presence of Cepheid variables (Caldwell et
al. 1988).  Some of the reddest stars with $V-I > 1.5$ overlap the
older isochrones with $\log t =9.0$ to $10.0$ yr, showing that old
stars are also present.

This interpretation depends on the adopted distance which cannot be
uniquely obtained from the isochrones alone; the isochrones can be
shifted upward without invalidating the fit. The additional data
discussed above is needed to constrain the possible distance range.

The reddest stars reach $V-I=2.5$, bluer than the extremes predicted
by the $Z=0.004$ isochrones ($V-I=3.4$). There are a few red stars
that are redder than the $Z=0.001$ isochrones would predict,
indicating that $0.001<Z<0.004$.  Likewise, the LMC supergiant branch
is more extended to the red than the IC~5152 supergiant branch. {}From
this comparison with the isochrones and with the color-magnitude
diagrams of similar galaxies, we conclude that the intermediate-age
stellar component of the IC~5152 disk has a metallicity $Z=0.002$.

The central region of this galaxy was observed with the Faint Object
Camera on the Hubble Space Telescope in the ultraviolet.  We have
retrieved the FOC image from the HST archives, in order to make a
comparison with our optical images of the same region. This image
taken with the F220W filter is centered at
$\alpha_{2000}=22:02:41.90$, $\delta_{2000}=-51:17:44.0$, and covers
$22\times22$ arcsec. It shows obvious star-forming activity, in the
form of numerous ultraviolet point sources, which are probably O and B-type
stars (Maoz et al. 1996). There is no obvious bright nucleus present
in this field. Figure 12 shows the $V$ and $I$-band NTT images of the
central $30\times30$ arcsec, for comparison with plate 1 of Maoz et
al.  (1996). Most of the ultraviolet sources can be identified with optical
sources seen in the $V$-band image. The sources in common have blue
$V-I$ colors, and some are clearly blends of 2--3 stars due to the
lower resolution of the NTT images compared with the HST ones.  

The brightest and reddest source in Figure 12 (located at $X=1707.98,~
Y=599.31$, with $I=18.12$, and $V-I=1.65$) has a faint counterpart in
the FOC image.  If it is an unreddened single red supergiant, it would
have $M_{bol} = -7.5$, but it should not show any ultraviolet
counterpart.  The ultraviolet detection indicates it could be a
compact stellar cluster. This source is located in the geometric
center of IC5152 judging from the POSS plates, and we suggest it could
be the highly reddened nucleus of this galaxy.

\section{Other Population Tracers}

\subsection{Star Clusters in IC 5152}

The status of globular cluster research in galaxies of the Local Group
is summarized by Olszewski (1994).  IC 5152 does not have any globular
clusters known.  We have searched for globulars clusters within a
defined magnitude and color range, looking for objects that are nearly
resolved (i.e. $FWHM \geq FWHM_{psf}$).  For $m-M = 26.15$, and
$E(V-I) = 0$, the magnitude and color range of possible globulars was
taken to be $22>I>16.5$, and $0.7 \leq V-I \leq 1.2$, respectively,
from Reed (1985) and Harris (1996).  Even though the resolution is not
good enough to allow to discriminate the presence of fainter clusters
from chance overlapping of individual unrelated stars, we find 10
objects brighter than $I \approx 20$ which have color similar to
Galactic globulars.  Based on the field color-magnitude diagram shown
in Figure 6, we would have expected 4 such objects (although the field
itself can contain IC~5152 globulars if the cluster system of this
galaxy is extended). Of the 10 candidates, 6 have colors and
magnitudes consistent with blends between a bright main sequence star
and a bright red supergiant star. The rest, however, cannot be
explained in this way because they are too bright.

\subsection{Carbon Stars}

No systematic search for C stars has been made in IC~5152.  The C star
luminosity functions given by Brewer et al. (1995) for M31, and by
Richer  (1981) for the LMC bar West field peak at $M_I = -4.4$,
with a total range of about $-3.4 \leq I_0 \leq -5.0$.  C stars are
expected to be very red, with $2<V-I<3$.  These expected magnitudes
and colors are coincident with the location of the reddest stars seen
in the lower portion of the color-magnitude diagram of IC~5152 (Figure
5).  Since these stars are located within the main body of the galaxy,
we conclude that there is a substantial number of candidate C stars in
this galaxy.

It is interesting to note that IC 5152 was detected by IRAS (source
identification IRAS21594-51).  Saunders et al. (1990) measured a total
luminosity at $60 \mu m$ of $\log L_{60}/L_{\odot} = 7.48$ for this
galaxy, adopting a distance $D=1.5$ Mpc.  We suggest that part of this
emission (of order 1 Jy at 12 micron) could be due to circumstellar
shells from mass-losing carbon stars in this galaxy.  Most
of the IRAS flux may however be due to the HII regions, as has been
found for the dIr galaxy NGC~6822 (Israel et al. 1996).

\subsection{HII Regions}

Being a late type galaxy, Sdm IV-V, IC~5152 is reasonably active at
forming stars at the present time (e.g. Elmegreen et al 1994). {}From
the main sequence (Figure 11), we see that the youngest stars are $\sim
10^7$ yr old indicative of on-going star formation.  The HII regions
of IC 5152 have been studied by Talent (1980) and by Webster et
al. (1983).  The former data is not published but the result is quoted
(Skillman et al. 1989, Marconi et al. 1994, Lisenfeld \&\ Ferrara
1998) as $12+\log(O/H)=8.36$. Webster et al. derive $12+\log(O/H)=8.4$
for a single HII region.  However, Hidalgo-G\'amez \& Olofsson (1998)
derive a rather lower abundance of $12+\log(O/H)=7.71$ based on the
data of Webster et al.  

The color-magnitude diagram indicates that the stellar component of
the IC~5152 disk has a metallicity $Z=0.002$, in approximate agreement
with the oxygen abundances listed by Hidalgo-G\'amez \& Olofsson
(1998). It is reasonable to assume that young supergiants and HII
regions will have the same metallicity. Galaxies with a relatively
young population are less likely to show a high value of [O/Fe]
(e.g. Walsh et al.  1997) so that we find better consistency with
Hidalgo-G\'amez \& Olofsson (1998).

Skillman et al. (1989) point out that IC~5152 falls well away from the
relation defined by other dwarf galaxies in the $12+\log(O/H)$ $vs$
$M_B$ plane. This galaxy is the most discrepant point in their sample,
being too faint and/or too metal-rich. The more accurate distance
presented here somewhat alleviates this problem, but not enough to
bring IC~5152 into agreement with the metallicity-luminosity relation
followed by other dwarfs. However, the discrepancy disappears when
using the lower abundance published by Hidalgo-G\'amez \& Olofsson
(1998), which is supported by the metallicity of the stellar component
in IC~5152.  (Note that Lisenfeld \&\ Ferraro (1998) use an extreme
distance of 3.0 Mpc, and Hidalgo-G\'amez \& Olofsson (1998) use 2.3
Mpc, shifting IC 5152 off the relation in both studies.)

Further observations would be desirable, since the preferred value for
the oxygen abundance is based on a single HII region for which two
discrepant abundance determinations exist. The possibility that recent
enrichment has caused metallicity variations within the gaseous disk
can also not be ruled out. At the smaller distances of 300 and 760
kpc, the galaxy would fall far above the expected relation even for
the reduced oxygen abundance, which is one of the arguments to prefer
the larger distance of 1.7 Mpc.

\subsection{Planetary Nebulae}

No planetary nebulae  in IC 5152 have been reported in the
literature. Given the luminosity of this galaxy $M_B = -14.8$, we
expect about ten planetary nebulae  to be present. 
If  planetary nebulae could be found, they would
allow for accurate oxygen-abundance determinations for the
intermediate-age stellar population.

\subsection{Cepheid Variables}

The only systematic search for Cepheid variables in IC~5152 reported
in the literature is that of Caldwell et al. (1988).  The distance
modulus inferred from these data is $m-M_0=26.0$ (see van den Bergh
1994a), corresponding to $D=1.6$ Mpc, in agreement with the distance
obtained in the present work. 

\subsection{The Gaseous Component: HI Observations}

Dwarf galaxies have dark matter halos, their total $M/L$ being larger
than that of normal spiral galaxies (e.g. Carignan \& Freeman 1988,
C\^ot\'e 1995), and IC~5152 seems to be no exception.  HI measurements
for IC 5152 are given by Huchtmeier \& Richter (1986).  They list a
velocity of $V_{LG}=+53$, a diameter of $A=1.95$ kpc (or $A=2.1$ kpc
using our revised distance), and infer $M_T/L_B=3.5$.

\section{Galaxy group association}

Our main result is that the distance of $D=1.7$ Mpc would place IC~5152
outside the accepted boundaries of the Local Group.  The distance,
together with the small positive velocity with respect to the Local
Group, implies that this galaxy is just turning around from the Hubble
flow (Peebles 1995).  

In the $\cos \theta$ versus velocity diagram, as commented by van den
Bergh (1994b), the position of IC5152 is only barely outside the
region occupied by most certain members of the Local Group. Thus, its
membership is still uncertain even with the improved distance
estimate.  IC~5152 is at a distance of 1.6 Mpc from the center of the
Local Group, at right angles to the direction of M31. Using a mass of
the Local group of $4 \times 10^{12}\, M_\odot$ indicates that its
radial velocity is less than the escape velocity, however the
tangential velocity is not known and may be dominant: a circular orbit
would require approximately $v_t=100\,$km/s. 

The mass distribution outside the Local Group also plays a role in the
membership problem.  IC 5152 is in fact closer to NGC 300 in the
Sculptor Group (1.2 Mpc) than to the main galaxies in the Local Group.
The Sculptor Group has been found to have a large population
of dwarf galaxies (C\^ot\'e et al. 1997), although their survey did
not cover the area of IC 5152.  The result of C\^ot\'e et al. (1997)
shows that the catalogues of dwarf galaxies beyond 1 Mpc is still very
incomplete. It is therefore possible that more such isolated dwarfs in
between the two groups exist, and even a common halo of dwarfs
could be envisaged.

There are no other known galaxies in this direction at this
approximate distance. The dwarfs WLM, Tucana, and NGC~1613 are in the
same quadrant, but are much closer than IC~5152, being real members of
the Local Group (van den Bergh 1994b).  IC 5152 appears to be one of
the most isolated nearby dwarf galaxies known, located in between the
Sculptor and Local Groups but distant enough to have so far evolved
without interaction with either of these groups.  The question arises
what drives the star formation in such an isolated galaxy. The
presence of numerous UV-bright stars in the central regions suggests a
relatively high rate of star formation, showing that star formation in
dwarf galaxies does not require external disturbances.  Van den Bergh
(1994b) has suggested that for the smallest dwarf galaxies
($M_B>-14.0$) the star formation history depends on the distance to a
large galaxy, with the nearest galaxies having most of their star
formation at an early phase and more isolated dwarfs have a higher
fraction of recent star formation. IC 5152 may form an extreme case of
the latter.

\section{Conclusions}

We have presented VI photometry for $553$ stars in the field of the
dwarf irregular galaxy IC 5152, a suspect member of the Local Group,
and have made the first CCD color-magnitude diagrams and luminosity
functions obtained for this galaxy.

By direct comparison of the VI color--magnitude diagram with those of
similar galaxies, we derive a distance modulus of $m-M_0=26.15 \pm
0.2$ for this galaxy. The distance of $D=1.7$ Mpc places IC~5152
just outside the accepted boundaries of the Local Group, leaving its
membership open.

Based on the color-magnitude diagrams, we have explored the stellar
population content, metallicities and ages for the main body of IC
5152.  The main sequence suggests that significant numbers of young
stars are present in this galaxy, and the extent of the supergiant and
asymptotic giant branches suggests the presence of stars with
intermediate-age with metallicity $Z\approx 0.002$, lower than
previously suggested.  The present result places the galaxy on the
luminosity--metallicity sequence of dwarf galaxies of Skillman et al.
(1989).  The central region is an active site of star formation, as
found by comparing our optical photometry with Hubble Space Telescope
ultraviolet images. On the basis of the ultraviolet and optical images
we identify a candidate for a potential nucleus.  A few globular
cluster candidates are found, which need to be confirmed
spectroscopically.

We also construct deep I-band luminosity functions, and a bolometric
luminosity function for the red stellar component of this galaxy.
Based on these we confirm the presence of an intermediate age stellar
population of AGB stars, and estimate that the brightest red
supergiants reach $M_{bol}=-9.5$ mag.

\acknowledgements{ We are grateful for the remote-control operators
J. Rodriguez and V.  Reyes, and the telescope operator M. Pizarro, for
their efficiency and expertise at the NTT.  This work was performed in
part under the auspices of the U. S. Department of Energy by Lawrence
Livermore National Laboratory under Contract W-7405-Eng-48. }

\clearpage

\begin{figure}
\caption{
Image  of IC 5152 from the POSS. North is up, and East to the left, with
a scale of 20$\times 20$ arcmin.  The field studied in this paper is indicated
by the rectangle. At the distance of IC 5152,
$1\arcmin = 0.5 ~kpc$. Note the spikes of the bright star HD209142, with 
$V=7.9$,
overlapped on the NW corner of this galaxy. }
\end{figure}

\begin{figure}
\caption{
Portion of the 
V image  of IC 5152 taken at the NTT. North is up, and East to the left, with
a field of about $3'\times 3'$, equivalent to about 1.5 kpc on a side.}
\end{figure}

\begin{figure}
\caption{
Map of all the stars detected in the field
of IC 5152 taken with the NTT. The sizes of the circles are proportional
to the I-band magnitudes of the stars.
The scale is $0.268$ arcsec/pixel, and
North is up, and East is to the left.
}
\end{figure}

\begin{figure}
\caption{
DAOPHOT  $V$ and $I$-band errors as function of magnitude.
}
\end{figure}

\begin{figure}
\caption{
Color-magnitude diagrams for all the stars with $V$ and $I$ measured in IC 
5152.
}
\end{figure}

\begin{figure}
\caption{
Color-magnitude diagrams of the background IC 5152 field.
}
\end{figure}

\begin{figure}
\caption{
I- band luminosity function including all the stars with photometry
in the $6.7$ arcmin$^2$ field of the IC 5152 disk.  
The supergiant branch termination is seen at about $I = 18.5$,
and the AGB termination at about $I = 21.0$.
}
\end{figure}

\begin{figure}
\caption{
I band luminosity function of the IC 5152 background field. This field
covers $16.5$ arcmin$^2$.
}
\end{figure}

\begin{figure}
\caption{
Bolometric luminosity function of IC 5152, including all the red stars 
with $V-I>0.8$.  The brightest red supergiants in this galaxy reach
$M_{bol}=-10.5$, and the supergiant branch break is seen at $M_{bol}=-8$.
}
\end{figure}

\begin{figure}
\caption{
Comparison of IC5152  with the color-magnitude diagrams of
the dIrr galaxies NGC3109 and WLM.
}
\end{figure}

\begin{figure}
\caption{
VI color-magnitude diagram for IC 5152 compared with the isochrones
of Bertelli et al. (1994) for $Z=0.001$ (left panel) and $Z=0.004$ (right 
panel).
The isochrones have been shifted according to  $m-M=26.15$ and 
$E_{V-I}=0$ measured in this work.
{}From top to bottom, the isochrones shown have ages $\log t= 6.6, 7.0, 7.8, 
8.0,
8.7, 9.0, 9.7$
and $10.0$ years.
}
\end{figure}

\begin{figure}
\caption{
Central I-band (left panel) and V-band (right panel) images of IC5152
covering $30\times30$ arcsec. Compare these images with Plate 1 of
Maoz et al. (1996). North is up, East to the left. Most of the sources
seen in the V-band image have ultraviolet counterparts, indicating that this
region has active star formation. The brightest source
in the I-band image is very faint in the ultraviolet, and 
could be the highly reddened nucleus of this galaxy.
}
\end{figure}

\begin{planotable}{lrllll}
\small
\footnotesize
\tablewidth{0pt}
\scriptsize
\tablecaption{Summary of parameters for IC 5152}
\tablehead{
\multicolumn{1}{c}{Parameter}&
\multicolumn{1}{c}{Value}&
\multicolumn{1}{c}{}&
\multicolumn{1}{c}{Reference}}
\startdata
$(\alpha, ~\delta_{2000})$&$22:02:41.22, ~-51:17:47.9$& & 4 \nl
$(l, ~b)$                &$343.91^{\circ}, ~-50.19^{\circ}$& & 4 \nl
Distance modulus         &$26.15\pm 0.20$ &    & 1 \nl
Distance                 &$1.7 ~Mpc$&          & 1 \nl
$Type$                   &  Sdm IV-V &         & 4 \nl
$M_B$                    &  $-14.8$  &         & 2 \nl
$E(V-I)$                 &   0.00   &          & 1 \nl
b/a$_{disk}$             &   0.6    &          & 4 \nl
$V_{LG}$                 &   +53 km s$^{-1}$ & & 2 \nl
Inclination              &   $55^{\circ}$   &  & 2 \nl
Diameter                 &   2.1 kpc    &      & 2 \nl
$Z_{disk}$               &   0.002  &          & 3 \nl
\enddata
\tablerefs{ \\
1. This work. \\
2. Huchtmeier \& Richter 1986. \\
3. Skillman et al. 1989a,b. \\
4. Sandage \& Bedke 1985. \\
}
\end{planotable}

\begin{planotable}{rrrrrrrrrrrrrr}
\small
\footnotesize
\tablewidth{0pt}
\scriptsize
\tablecaption{Photometry of IC~5152}
\tablehead{
\multicolumn{1}{c}{X}&
\multicolumn{1}{c}{Y}&
\multicolumn{1}{c}{V}&
\multicolumn{1}{c}{$\sigma_V$}&
\multicolumn{1}{c}{I}&
\multicolumn{1}{c}{$\sigma_I$}&
\multicolumn{1}{c}{$V-I$}&
\multicolumn{1}{c}{X}&
\multicolumn{1}{c}{Y}&
\multicolumn{1}{c}{V}&
\multicolumn{1}{c}{$\sigma_V$}&
\multicolumn{1}{c}{I}&
\multicolumn{1}{c}{$\sigma_I$}&
\multicolumn{1}{c}{$V-I$}}
\startdata
   1136.38& 633.89& 22.32& 0.09& 21.28& 0.06& 1.04\ & 1140.38& 291.20& 18.75& 
0.00& 16.53& 0.00& 2.22\nl
   1149.74& 172.46& 23.04& 0.16& 22.45& 0.19& 0.58\ & 1156.24& 544.71& 23.03& 
0.19& 23.51& 0.50& -0.48\nl
   1169.46& 133.12& 20.51& 0.02& 18.41& 0.01& 2.10\ & 1178.83& 190.40& 23.30& 
0.24& 21.55& 0.09& 1.75\nl
   1193.98& 303.99& 23.32& 0.26& 23.01& 0.29& 0.31\ & 1196.11& 539.47& 23.79& 
0.39& 22.73& 0.24& 1.06\nl
   1197.83& 612.09& 22.48& 0.11& 21.50& 0.08& 0.98\ & 1200.12& 327.34& 22.56& 
0.13& 20.29& 0.02& 2.27\nl
   1203.06& 507.52& 21.55& 0.05& 21.62& 0.08& -0.07\ & 1204.22& 636.65& 23.81& 
0.45& 22.42& 0.15& 1.39\nl
   1205.33& 651.59& 24.16& 0.53& 21.91& 0.10& 2.25\ & 1212.77& 639.74& 21.78& 
0.05& 22.02& 0.11& -0.24\nl
   1217.85& 173.72& 23.85& 0.38& 23.30& 0.43& 0.56\ & 1226.08& 230.57& 24.20& 
0.47& 22.44& 0.18& 1.76\nl
   1226.28& 479.12& 22.53& 0.09& 21.16& 0.05& 1.38\ & 1229.38& 615.15& 22.87& 
0.16& 22.06& 0.12& 0.80\nl
   1229.96& 319.60& 22.29& 0.09& 21.41& 0.06& 0.88\ & 1231.03& 345.36& 23.96& 
0.33& 23.67& 0.43& 0.29\nl
   1235.07& 513.31& 23.89& 0.30& 23.22& 0.43& 0.67\ & 1239.74& 627.15& 23.07& 
0.17& 21.64& 0.09& 1.43\nl
   1247.31& 261.34& 21.82& 0.06& 21.83& 0.12& -0.01\ & 1247.62& 516.54& 23.85& 
0.36& 22.23& 0.14& 1.63\nl
   1247.64& 351.17& 23.80& 0.30& 22.79& 0.25& 1.01\ & 1253.48& 556.40& 21.11& 
0.03& 20.51& 0.04& 0.60\nl
   1254.62& 300.28& 22.61& 0.12& 22.63& 0.22& -0.02\ & 1254.71& 334.65& 22.86& 
0.13& 21.81& 0.08& 1.06\nl
   1255.75& 504.02& 23.69& 0.30& 22.07& 0.14& 1.62\ & 1259.59& 320.79& 21.66& 
0.05& 21.98& 0.14& -0.32\nl
   1260.12& 505.51& 21.33& 0.03& 21.44& 0.07& -0.12\ & 1261.71& 146.11& 17.83& 
0.00& 16.78& 0.00& 1.05\nl
   1262.23& 427.45& 23.89& 0.34& 22.52& 0.18& 1.38\ & 1263.91& 491.22& 20.80& 
0.03& 19.29& 0.01& 1.51\nl
   1268.90& 612.41& 23.56& 0.27& 22.39& 0.14& 1.17\ & 1273.44& 252.86& 23.57& 
0.29& 23.17& 0.38& 0.40\nl
   1278.94& 452.42& 23.20& 0.21& 22.15& 0.13& 1.05\ & 1281.66& 553.17& 22.85& 
0.13& 21.98& 0.12& 0.87\nl
   1283.75& 630.51& 22.83& 0.13& 21.86& 0.10& 0.96\ & 1292.02& 612.76& 22.60& 
0.13& 21.76& 0.09& 0.84\nl
   1292.49& 338.51& 21.89& 0.05& 20.63& 0.04& 1.26\ & 1293.61& 554.10& 20.07& 
0.01& 18.53& 0.01& 1.54\nl
   1293.78& 645.11& 23.08& 0.18& 22.31& 0.36& 0.77\ & 1295.09& 618.43& 22.21& 
0.09& 21.94& 0.14& 0.27\nl
   1297.94& 507.24& 20.98& 0.08& 21.45& 0.09& -0.48\ & 1299.77& 521.89& 22.26& 
0.13& 22.62& 0.22& -0.36\nl
   1301.39& 495.49& 22.72& 0.16& 22.61& 0.20& 0.10\ & 1301.84& 506.78& 21.23& 
0.11& 21.51& 0.08& -0.28\nl
   1302.74& 237.53& 21.70& 0.05& 21.37& 0.06& 0.33\ & 1308.38& 579.52& 22.59& 
0.13& 21.98& 0.12& 0.62\nl
   1308.51& 500.39& 21.71& 0.08& 22.17& 0.17& -0.45\ & 1308.92& 193.84& 21.90& 
0.06& 21.86& 0.10& 0.04\nl
   1312.48& 585.86& 22.03& 0.07& 22.18& 0.16& -0.15\ & 1313.47& 663.85& 22.37& 
0.08& 20.81& 0.04& 1.56\nl
   1314.00& 265.23& 21.11& 0.03& 21.48& 0.07& -0.37\ & 1317.29& 227.35& 20.92& 
0.02& 19.30& 0.01& 1.62\nl
   1317.85& 466.60& 23.54& 0.28& 22.75& 0.28& 0.78\ & 1322.90& 561.67& 20.62& 
0.02& 18.91& 0.01& 1.72\nl
   1327.33& 360.35& 23.52& 0.24& 22.80& 0.22& 0.73\ & 1327.54& 490.28& 22.82& 
0.15& 22.10& 0.14& 0.72\nl
   1327.94& 532.99& 23.34& 0.21& 21.74& 0.09& 1.60\ & 1329.80& 457.31& 20.53& 
0.02& 20.73& 0.03& -0.20\nl
   1330.47& 582.96& 22.60& 0.15& 21.87& 0.11& 0.73\ & 1331.10& 389.03& 21.11& 
0.04& 21.75& 0.09& -0.64\nl
   1332.79& 396.99& 22.41& 0.09& 22.63& 0.22& -0.23\ & 1335.57& 487.26& 23.27& 
0.23& 21.91& 0.12& 1.37\nl
   1338.68& 550.17& 22.10& 0.08& 19.88& 0.02& 2.22\ & 1340.76& 532.52& 22.66& 
0.13& 22.51& 0.19& 0.15\nl
   1341.32& 376.16& 22.18& 0.09& 21.62& 0.09& 0.56\ & 1343.29& 606.80& 24.02& 
0.50& 22.27& 0.17& 1.76\nl
   1343.48& 392.84& 20.77& 0.05& 21.21& 0.06& -0.44\ & 1344.63& 420.05& 22.64& 
0.12& 23.02& 0.28& -0.39\nl
   1346.61& 541.23& 20.71& 0.03& 20.82& 0.04& -0.11\ & 1348.74& 554.62& 22.51& 
0.11& 22.57& 0.23& -0.07\nl
   1350.00& 441.29& 21.44& 0.03& 21.71& 0.10& -0.27\ & 1350.44& 622.25& 22.46& 
0.12& 22.31& 0.24& 0.15\nl
   1351.51& 521.07& 21.97& 0.08& 21.05& 0.05& 0.92\ & 1352.04& 560.85& 21.70& 
0.07& 21.41& 0.07& 0.28\nl
   1354.80& 511.74& 22.45& 0.12& 21.39& 0.07& 1.07\ & 1354.81& 602.23& 23.31& 
0.22& 22.74& 0.29& 0.57\nl
   1356.98& 525.62& 20.89& 0.03& 21.18& 0.06& -0.29\ & 1358.19& 474.07& 22.21& 
0.08& 21.18& 0.06& 1.03\nl
   1359.51& 250.21& 22.87& 0.14& 22.21& 0.13& 0.66\ & 1362.96& 518.61& 20.87& 
0.04& 21.40& 0.08& -0.53\nl
   1364.68& 556.17& 22.20& 0.10& 22.29& 0.16& -0.09\ & 1366.29& 355.02& 24.16& 
0.48& 23.15& 0.33& 1.01\nl
   1366.67& 488.10& 22.08& 0.08& 21.58& 0.09& 0.50\ & 1368.02& 199.62& 20.52& 
0.02& 19.75& 0.01& 0.76\nl
   1369.11& 504.07& 19.76& 0.03& 19.96& 0.03& -0.20\ & 1372.36& 426.11& 21.54& 
0.05& 20.29& 0.03& 1.25\nl
   1376.12& 546.94& 21.80& 0.05& 21.31& 0.07& 0.48\ & 1376.72& 449.13& 23.10& 
0.24& 21.31& 0.05& 1.79\nl
   1377.96& 592.86& 23.08& 0.17& 21.52& 0.08& 1.55\ & 1378.21& 540.29& 20.76& 
0.03& 20.76& 0.06& 0.00\nl
   1379.96& 363.86& 23.86& 0.39& 22.80& 0.28& 1.06\ & 1382.19& 576.75& 20.58& 
0.02& 20.53& 0.05& 0.05\nl
   1385.16& 413.52& 22.53& 0.12& 21.68& 0.08& 0.84\ & 1386.40& 321.56& 22.28& 
0.07& 22.15& 0.14& 0.13\nl
   1386.49& 586.57& 21.39& 0.04& 21.92& 0.10& -0.53\ & 1386.51& 547.41& 22.39& 
0.11& 23.19& 0.43& -0.81\nl
   1386.84& 522.92& 23.08& 0.21& 21.58& 0.08& 1.50\ & 1388.59& 606.20& 22.96& 
0.20& 22.33& 0.15& 0.64\nl
   1389.25& 648.62& 23.40& 0.24& 21.77& 0.10& 1.63\ & 1394.07& 116.20& 22.21& 
0.08& 21.48& 0.09& 0.73\nl
   1395.98& 552.49& 20.47& 0.03& 20.49& 0.04& -0.01\ & 1397.66& 412.18& 23.51& 
0.25& 21.61& 0.08& 1.91\nl
   1399.49& 542.57& 22.77& 0.14& 21.00& 0.06& 1.76\ & 1399.64& 579.27& 21.02& 
0.03& 20.84& 0.05& 0.18\nl
   1406.38& 535.16& 21.78& 0.06& 22.16& 0.15& -0.38\ & 1407.48& 473.06& 20.99& 
0.03& 21.21& 0.06& -0.22\nl
   1407.87& 569.76& 21.33& 0.04& 19.39& 0.01& 1.95\ & 1409.11& 451.51& 20.00& 
0.01& 18.22& 0.01& 1.78\nl
   1409.41& 477.50& 20.90& 0.03& 21.41& 0.06& -0.51\ & 1409.73& 504.57& 22.27& 
0.09& 21.32& 0.06& 0.95\nl
   1413.23& 429.83& 22.63& 0.13& 21.47& 0.07& 1.16\ & 1414.53& 465.01& 21.99& 
0.08& 21.96& 0.11& 0.03\nl
   1416.10& 620.47& 21.61& 0.05& 20.41& 0.03& 1.20\ & 1416.66& 449.69& 22.70& 
0.14& 22.74& 0.21& -0.04\nl
   1418.23& 532.64& 23.72& 0.39& 22.89& 0.29& 0.83\ & 1418.35& 602.06& 23.18& 
0.22& 22.86& 0.29& 0.32\nl
   1422.73& 612.17& 21.51& 0.04& 20.85& 0.05& 0.66\ & 1423.70& 284.34& 22.75& 
0.15& 21.79& 0.10& 0.97\nl
   1424.18& 539.87& 24.22& 0.49& 21.94& 0.13& 2.28\ & 1424.52& 454.57& 23.40& 
0.23& 22.18& 0.15& 1.22\nl
   1424.66& 618.28& 21.80& 0.05& 20.75& 0.04& 1.05\ & 1426.39& 439.47& 22.06& 
0.10& 23.35& 0.52& -1.29\nl
   1428.91& 576.78& 23.92& 0.48& 21.89& 0.11& 2.03\ & 1429.35& 480.43& 21.09& 
0.04& 19.78& 0.02& 1.31\nl
   1429.51& 528.12& 22.20& 0.10& 21.00& 0.05& 1.21\ & 1429.92& 565.37& 23.02& 
0.16& 22.02& 0.14& 1.00\nl
   1430.03& 317.72& 23.01& 0.16& 21.73& 0.09& 1.28\ & 1430.89& 590.26& 22.22& 
0.09& 20.97& 0.05& 1.25\nl
   1431.07& 100.77& 24.09& 0.39& 22.44& 0.21& 1.65\ & 1431.66& 545.29& 22.06& 
0.08& 21.13& 0.06& 0.94\nl
   1433.15& 423.54& 21.17& 0.04& 19.75& 0.03& 1.43\ & 1433.58& 576.85& 22.95& 
0.16& 22.29& 0.19& 0.66\nl
   1436.19& 652.20& 21.35& 0.04& 21.49& 0.09& -0.14\ & 1437.00& 456.81& 21.44& 
0.05& 20.41& 0.04& 1.03\nl
   1438.90& 585.88& 23.04& 0.18& 22.06& 0.14& 0.98\ & 1439.30& 431.96& 21.24& 
0.05& 21.51& 0.09& -0.27\nl
   1439.93& 634.17& 23.33& 0.25& 21.69& 0.11& 1.64\ & 1440.60& 338.44& 21.91& 
0.07& 22.04& 0.14& -0.13\nl
   1441.53& 220.43& 23.95& 0.46& 21.87& 0.09& 2.09\ & 1442.07& 501.87& 21.52& 
0.06& 20.41& 0.04& 1.11\nl
   1443.16& 486.49& 23.21& 0.18& 22.43& 0.18& 0.77\ & 1444.16& 474.48& 22.05& 
0.07& 21.83& 0.11& 0.21\nl
   1444.24& 608.35& 23.98& 0.38& 22.78& 0.23& 1.19\ & 1444.96& 629.56& 23.75& 
0.30& 22.00& 0.12& 1.76\nl
   1449.93& 441.43& 22.39& 0.11& 22.26& 0.14& 0.13\ & 1450.87& 557.13& 23.74& 
0.38& 21.92& 0.12& 1.82\nl
   1451.41& 460.05& 21.55& 0.05& 20.88& 0.05& 0.67\ & 1452.09& 509.79& 23.41& 
0.29& 21.70& 0.10& 1.71\nl
   1452.15& 568.30& 20.72& 0.07& 21.11& 0.09& -0.40\ & 1453.08& 564.59& 21.31& 
0.07& 21.47& 0.12& -0.15\nl
   1453.30& 585.40& 22.71& 0.13& 21.95& 0.12& 0.76\ & 1454.49& 467.31& 21.37& 
0.06& 20.91& 0.06& 0.46\nl
   1455.43& 615.92& 23.17& 0.20& 21.77& 0.08& 1.39\ & 1455.68& 340.80& 21.81& 
0.05& 20.71& 0.03& 1.10\nl
   1456.82& 478.48& 22.17& 0.08& 21.14& 0.06& 1.04\ & 1456.92& 512.68& 22.41& 
0.13& 21.95& 0.11& 0.46\nl
   1457.65& 625.79& 23.66& 0.28& 22.07& 0.16& 1.58\ & 1457.76& 572.82& 21.41& 
0.07& 21.86& 0.11& -0.45\nl
   1458.23& 461.08& 21.40& 0.03& 21.54& 0.08& -0.14\ & 1458.58& 591.35& 20.65& 
0.02& 20.88& 0.05& -0.24\nl
   1458.84& 526.12& 21.68& 0.07& 21.40& 0.08& 0.28\ & 1459.38& 424.86& 22.40& 
0.09& 22.04& 0.12& 0.36\nl
   1460.39& 565.26& 21.12& 0.07& 21.77& 0.12& -0.65\ & 1463.77& 587.20& 22.97& 
0.18& 21.48& 0.08& 1.49\nl
   1464.02& 471.59& 22.01& 0.06& 21.57& 0.08& 0.44\ & 1464.81& 373.69& 23.01& 
0.17& 22.01& 0.13& 1.00\nl
   1465.57& 315.54& 22.77& 0.14& 21.00& 0.07& 1.77\ & 1465.80& 582.19& 22.30& 
0.11& 23.26& 0.38& -0.96\nl
   1467.88& 564.29& 22.60& 0.13& 23.15& 0.38& -0.54\ & 1468.09& 425.38& 23.92& 
0.44& 22.90& 0.24& 1.02\nl
   1468.34& 515.15& 21.70& 0.07& 20.48& 0.03& 1.22\ & 1469.55& 459.70& 22.13& 
0.09& 21.35& 0.08& 0.78\nl
   1470.14& 388.94& 21.84& 0.05& 20.73& 0.04& 1.10\ & 1470.50& 593.74& 20.82& 
0.04& 21.13& 0.05& -0.31\nl
   1471.74& 501.92& 21.15& 0.05& 21.21& 0.08& -0.06\ & 1473.98& 559.52& 22.55& 
0.14& 22.46& 0.21& 0.09\nl
   1474.31& 468.67& 22.69& 0.15& 23.18& 0.41& -0.49\ & 1479.60& 481.52& 22.75& 
0.14& 21.56& 0.10& 1.20\nl
   1479.74& 544.11& 20.79& 0.03& 19.19& 0.02& 1.59\ & 1480.43& 646.13& 23.56& 
0.28& 23.13& 0.33& 0.43\nl
   1482.35& 552.31& 22.17& 0.09& 21.63& 0.09& 0.54\ & 1482.80& 510.07& 21.10& 
0.03& 21.62& 0.06& -0.52\nl
   1483.71& 606.17& 21.94& 0.08& 20.36& 0.04& 1.58\ & 1484.69& 461.55& 22.39& 
0.11& 21.45& 0.07& 0.94\nl
   1485.02& 525.32& 21.94& 0.07& 20.39& 0.03& 1.55\ & 1486.38& 448.12& 21.94& 
0.07& 20.95& 0.05& 0.99\nl
   1486.91& 508.80& 22.08& 0.07& 22.46& 0.18& -0.38\ & 1487.28& 467.53& 23.74& 
0.34& 21.75& 0.07& 1.99\nl
   1487.61& 561.43& 22.74& 0.18& 21.31& 0.08& 1.44\ & 1488.12& 332.34& 21.33& 
0.04& 19.87& 0.02& 1.46\nl
   1488.13& 307.07& 22.72& 0.13& 21.73& 0.10& 1.00\ & 1491.39& 529.13& 21.64& 
0.05& 21.39& 0.08& 0.26\nl
   1491.79& 306.07& 23.83& 0.36& 23.18& 0.40& 0.65\ & 1492.46& 597.67& 21.87& 
0.06& 20.73& 0.04& 1.14\nl
   1493.25& 492.37& 22.42& 0.10& 21.58& 0.07& 0.85\ & 1494.08& 569.02& 20.83& 
0.03& 21.39& 0.07& -0.56\nl
   1494.48& 575.44& 22.23& 0.10& 21.60& 0.09& 0.63\ & 1495.59& 537.26& 19.19& 
0.01& 19.35& 0.01& -0.16\nl
   1497.41& 483.03& 21.36& 0.05& 21.41& 0.08& -0.05\ & 1498.41& 506.89& 21.28& 
0.06& 20.03& 0.02& 1.25\nl
   1498.80& 633.88& 22.61& 0.17& 21.51& 0.09& 1.10\ & 1500.95& 402.88& 20.85& 
0.03& 19.05& 0.01& 1.80\nl
   1501.03& 587.38& 22.49& 0.09& 20.59& 0.04& 1.91\ & 1501.45& 619.68& 22.49& 
0.10& 21.99& 0.11& 0.50\nl
   1501.51& 640.76& 21.53& 0.05& 21.50& 0.09& 0.03\ & 1503.09& 559.14& 22.25& 
0.12& 20.83& 0.04& 1.42\nl
   1503.70& 604.17& 23.64& 0.33& 22.71& 0.29& 0.93\ & 1503.85& 581.22& 21.35& 
0.04& 21.27& 0.08& 0.08\nl
   1504.47& 576.76& 20.73& 0.03& 20.90& 0.07& -0.17\ & 1505.05& 476.60& 21.46& 
0.05& 21.53& 0.10& -0.07\nl
   1506.97& 351.73& 21.95& 0.05& 22.16& 0.13& -0.21\ & 1507.30& 610.18& 20.55& 
0.03& 20.76& 0.05& -0.22\nl
   1507.42& 630.99& 22.34& 0.11& 22.09& 0.17& 0.25\ & 1509.95& 592.72& 22.46& 
0.18& 21.12& 0.07& 1.34\nl
   1512.07& 654.91& 23.31& 0.28& 22.30& 0.16& 1.01\ & 1513.03& 540.19& 21.68& 
0.07& 21.57& 0.09& 0.12\nl
   1513.05& 522.82& 20.27& 0.02& 20.64& 0.03& -0.37\ & 1513.22& 504.59& 21.15& 
0.04& 21.12& 0.06& 0.04\nl
   1513.29& 439.18& 20.78& 0.02& 18.95& 0.01& 1.83\ & 1513.45& 574.20& 20.87& 
0.03& 20.17& 0.02& 0.70\nl
   1513.54& 622.13& 22.67& 0.16& 21.93& 0.12& 0.74\ & 1513.72& 559.31& 21.55& 
0.05& 20.40& 0.03& 1.15\nl
   1515.85& 353.85& 23.63& 0.35& 22.28& 0.17& 1.35\ & 1516.32& 602.72& 22.33& 
0.12& 21.57& 0.08& 0.76\nl
   1517.92& 567.95& 20.33& 0.02& 20.72& 0.04& -0.39\ & 1518.10& 619.55& 22.76& 
0.17& 21.97& 0.14& 0.80\nl
   1520.05& 527.86& 22.02& 0.08& 21.36& 0.07& 0.66\ & 1520.99& 380.90& 20.81& 
0.04& 19.59& 0.02& 1.22\nl
   1522.43& 583.62& 20.73& 0.04& 19.39& 0.02& 1.33\ & 1522.82& 600.36& 22.35& 
0.13& 21.31& 0.07& 1.04\nl
   1523.97& 589.16& 20.84& 0.04& 19.55& 0.02& 1.29\ & 1524.23& 468.05& 22.65& 
0.16& 22.47& 0.18& 0.18\nl
   1525.14& 452.23& 23.27& 0.25& 21.88& 0.09& 1.39\ & 1525.46& 433.07& 21.19& 
0.04& 21.77& 0.10& -0.59\nl
   1525.71& 342.81& 21.82& 0.06& 22.01& 0.10& -0.19\ & 1526.46& 298.60& 24.11& 
0.42& 22.40& 0.20& 1.71\nl
   1527.19& 476.67& 23.08& 0.16& 23.25& 0.43& -0.17\ & 1527.26& 507.16& 23.08& 
0.20& 21.77& 0.11& 1.32\nl
   1527.42& 103.43& 22.04& 0.07& 21.33& 0.07& 0.71\ & 1527.45& 654.78& 22.97& 
0.16& 22.95& 0.29& 0.02\nl
   1530.85& 637.47& 22.07& 0.10& 22.24& 0.17& -0.18\ & 1531.86& 327.75& 22.74& 
0.12& 21.47& 0.07& 1.26\nl
   1532.44& 600.76& 20.85& 0.03& 21.69& 0.11& -0.84\ & 1532.44& 556.84& 19.90& 
0.02& 20.03& 0.03& -0.13\nl
   1533.51& 651.45& 23.08& 0.18& 22.31& 0.17& 0.76\ & 1533.62& 520.26& 22.58& 
0.11& 21.17& 0.06& 1.41\nl
   1533.63& 536.89& 21.84& 0.09& 21.30& 0.10& 0.55\ & 1533.70& 351.19& 21.23& 
0.04& 19.95& 0.02& 1.28\nl
   1534.97& 250.15& 24.05& 0.49& 22.81& 0.36& 1.24\ & 1535.93& 388.85& 22.52& 
0.12& 21.85& 0.08& 0.67\nl
   1536.09& 560.72& 21.50& 0.05& 20.52& 0.05& 0.98\ & 1536.29& 440.23& 21.27& 
0.03& 21.15& 0.07& 0.12\nl
   1536.71& 196.99& 23.73& 0.33& 22.70& 0.22& 1.03\ & 1537.00& 396.15& 22.19& 
0.10& 21.02& 0.05& 1.17\nl
   1537.27& 530.37& 22.31& 0.11& 20.99& 0.05& 1.33\ & 1537.83& 461.99& 19.50& 
0.02& 19.74& 0.02& -0.24\nl
   1538.07& 521.18& 22.40& 0.10& 22.07& 0.15& 0.33\ & 1538.81& 274.53& 23.76& 
0.34& 22.78& 0.25& 0.98\nl
   1539.76& 485.97& 23.60& 0.28& 22.61& 0.18& 0.99\ & 1540.32& 588.02& 20.48& 
0.03& 19.25& 0.01& 1.22\nl
   1541.41& 559.19& 21.81& 0.08& 20.73& 0.04& 1.08\ & 1541.70& 625.85& 21.42& 
0.10& 20.36& 0.05& 1.06\nl
   1545.10& 249.33& 23.04& 0.16& 22.19& 0.12& 0.85\ & 1545.20& 562.64& 21.35& 
0.05& 20.64& 0.05& 0.71\nl
   1545.36& 499.45& 23.15& 0.22& 21.56& 0.08& 1.59\ & 1546.44& 386.76& 22.70& 
0.14& 21.53& 0.08& 1.17\nl
   1547.31& 364.45& 20.06& 0.01& 20.40& 0.03& -0.34\ & 1548.01& 575.77& 21.34& 
0.07& 20.46& 0.03& 0.88\nl
   1549.45& 447.65& 23.95& 0.38& 21.01& 0.06& 2.94\ & 1549.59& 531.88& 22.20& 
0.10& 21.16& 0.05& 1.04\nl
   1549.81& 615.46& 19.99& 0.02& 18.28& 0.01& 1.72\ & 1550.10& 518.12& 22.07& 
0.08& 22.30& 0.15& -0.22\nl
   1552.54& 655.25& 21.56& 0.09& 21.02& 0.05& 0.54\ & 1552.59& 575.29& 20.69& 
0.04& 20.89& 0.05& -0.20\nl
   1552.80& 266.37& 23.13& 0.22& 21.20& 0.05& 1.93\ & 1556.17& 370.36& 21.60& 
0.05& 21.99& 0.14& -0.40\nl
   1556.33& 363.94& 22.26& 0.11& 21.82& 0.14& 0.44\ & 1556.69& 572.72& 21.51& 
0.08& 20.94& 0.06& 0.57\nl
   1557.25& 384.33& 22.01& 0.08& 22.50& 0.21& -0.50\ & 1558.64& 601.44& 22.35& 
0.10& 21.07& 0.05& 1.27\nl
   1558.70& 397.71& 23.65& 0.36& 21.75& 0.11& 1.90\ & 1562.00& 333.64& 23.38& 
0.28& 22.76& 0.28& 0.62\nl
   1562.43& 449.13& 22.12& 0.09& 21.10& 0.05& 1.03\ & 1563.17& 606.46& 23.08& 
0.20& 21.22& 0.05& 1.86\nl
   1566.89& 514.48& 23.15& 0.28& 22.81& 0.32& 0.34\ & 1566.91& 377.24& 18.79& 
0.01& 17.77& 0.01& 1.01\nl
   1567.21& 307.70& 21.80& 0.07& 20.68& 0.03& 1.12\ & 1567.43& 535.59& 19.09& 
0.02& 19.00& 0.02& 0.09\nl
   1567.83& 612.05& 22.93& 0.18& 21.46& 0.07& 1.47\ & 1568.40& 402.70& 20.21& 
0.02& 20.11& 0.02& 0.09\nl
   1569.59& 555.76& 20.98& 0.05& 21.63& 0.11& -0.65\ & 1570.30& 582.62& 23.99& 
0.38& 22.85& 0.24& 1.14\nl
   1570.87& 608.71& 22.54& 0.12& 21.47& 0.09& 1.07\ & 1573.07& 145.02& 24.12& 
0.45& 23.21& 0.31& 0.91\nl
   1573.11& 525.97& 21.33& 0.07& 20.14& 0.03& 1.19\ & 1575.67& 447.51& 21.26& 
0.06& 20.91& 0.05& 0.35\nl
   1575.79& 359.57& 21.90& 0.06& 22.60& 0.22& -0.70\ & 1575.86& 559.40& 21.29& 
0.06& 22.61& 0.24& -1.32\nl
   1576.13& 612.99& 22.40& 0.12& 21.56& 0.08& 0.84\ & 1576.35& 629.50& 22.74& 
0.17& 22.78& 0.30& -0.05\nl
   1576.78& 427.21& 23.85& 0.41& 22.67& 0.21& 1.18\ & 1577.09& 553.87& 20.92& 
0.06& 21.90& 0.12& -0.98\nl
   1577.75& 647.70& 22.10& 0.10& 22.11& 0.14& 0.00\ & 1579.53& 450.69& 21.91& 
0.06& 21.97& 0.12& -0.06\nl
   1579.55& 497.26& 22.37& 0.10& 21.32& 0.06& 1.05\ & 1581.43& 619.18& 22.51& 
0.15& 22.58& 0.18& -0.07\nl
   1581.92& 345.89& 22.34& 0.10& 21.16& 0.07& 1.18\ & 1582.03& 612.74& 21.76& 
0.06& 21.99& 0.12& -0.23\nl
   1582.28& 330.64& 21.16& 0.03& 21.56& 0.08& -0.40\ & 1582.34& 636.22& 23.66& 
0.39& 21.35& 0.07& 2.31\nl
   1582.84& 406.49& 22.62& 0.15& 21.96& 0.11& 0.66\ & 1583.35& 377.23& 18.74& 
0.01& 17.52& 0.01& 1.22\nl
   1584.80& 663.21& 22.32& 0.10& 21.69& 0.09& 0.62\ & 1585.60& 577.96& 20.96& 
0.03& 19.43& 0.01& 1.53\nl
   1585.72& 278.22& 23.10& 0.18& 22.56& 0.22& 0.54\ & 1586.75& 202.28& 23.93& 
0.37& 23.03& 0.34& 0.90\nl
   1587.68& 545.85& 23.85& 0.45& 23.25& 0.46& 0.60\ & 1588.63& 520.26& 20.09& 
0.02& 20.60& 0.03& -0.51\nl
   1589.07& 431.30& 21.97& 0.12& 22.10& 0.14& -0.13\ & 1589.15& 612.32& 22.54& 
0.12& 21.68& 0.14& 0.86\nl
   1590.49& 482.68& 23.37& 0.27& 21.59& 0.09& 1.78\ & 1591.98& 663.64& 20.61& 
0.03& 21.00& 0.07& -0.39\nl
   1592.01& 637.43& 20.46& 0.02& 20.36& 0.04& 0.10\ & 1593.98& 531.43& 21.26& 
0.05& 21.21& 0.07& 0.05\nl
   1594.51& 420.43& 21.69& 0.05& 21.76& 0.09& -0.06\ & 1594.54& 470.30& 22.90& 
0.18& 22.38& 0.18& 0.53\nl
   1594.80& 425.30& 22.03& 0.10& 21.27& 0.07& 0.75\ & 1595.05& 619.38& 22.32& 
0.09& 21.74& 0.12& 0.58\nl
   1595.06& 373.15& 21.05& 0.02& 19.53& 0.01& 1.52\ & 1596.53& 560.55& 21.79& 
0.07& 21.60& 0.10& 0.19\nl
   1599.56& 453.86& 23.22& 0.21& 22.06& 0.13& 1.15\ & 1599.87& 509.28& 21.76& 
0.06& 20.45& 0.04& 1.31\nl
   1600.01& 400.45& 21.61& 0.07& 21.10& 0.05& 0.51\ & 1600.37& 377.27& 20.97& 
0.04& 21.17& 0.06& -0.20\nl
   1602.76& 351.07& 22.62& 0.12& 22.39& 0.19& 0.22\ & 1603.26& 364.07& 23.09& 
0.19& 21.97& 0.12& 1.12\nl
   1603.62& 220.94& 23.98& 0.44& 22.35& 0.16& 1.63\ & 1603.64& 592.04& 22.82& 
0.14& 21.00& 0.06& 1.82\nl
   1603.86& 597.08& 22.81& 0.17& 21.25& 0.08& 1.56\ & 1604.26& 537.99& 21.22& 
0.04& 19.70& 0.02& 1.53\nl
   1604.84& 641.89& 22.21& 0.10& 21.59& 0.08& 0.62\ & 1606.20& 269.23& 22.78& 
0.13& 22.85& 0.23& -0.07\nl
   1606.54& 389.72& 24.22& 0.51& 22.83& 0.26& 1.39\ & 1606.88& 296.33& 22.98& 
0.17& 21.56& 0.07& 1.42\nl
   1607.71& 620.31& 22.81& 0.24& 22.04& 0.12& 0.77\ & 1609.57& 499.15& 21.91& 
0.09& 21.97& 0.13& -0.07\nl
   1611.11& 614.26& 22.21& 0.09& 22.84& 0.25& -0.63\ & 1611.68& 510.12& 23.79& 
0.33& 22.36& 0.19& 1.43\nl
   1611.96& 551.92& 21.84& 0.10& 20.78& 0.06& 1.06\ & 1612.09& 453.03& 22.02& 
0.09& 21.65& 0.11& 0.38\nl
   1612.16& 290.34& 23.27& 0.28& 23.56& 0.53& -0.29\ & 1612.49& 531.53& 21.86& 
0.10& 22.50& 0.29& -0.64\nl
   1615.35& 445.72& 21.68& 0.06& 21.51& 0.08& 0.17\ & 1615.37& 582.79& 21.67& 
0.10& 21.94& 0.12& -0.27\nl
   1615.62& 556.55& 22.02& 0.11& 21.50& 0.09& 0.52\ & 1616.15& 338.26& 21.75& 
0.05& 22.18& 0.13& -0.43\nl
   1617.11& 401.68& 22.79& 0.14& 21.64& 0.10& 1.16\ & 1617.19& 473.39& 22.96& 
0.16& 22.14& 0.12& 0.82\nl
   1617.43& 542.29& 20.94& 0.03& 21.34& 0.09& -0.40\ & 1618.10& 413.48& 20.52& 
0.02& 20.72& 0.04& -0.21\nl
   1618.88& 640.99& 22.24& 0.10& 22.24& 0.14& 0.01\ & 1618.91& 569.03& 22.14& 
0.08& 22.23& 0.19& -0.09\nl
   1618.98& 612.63& 22.89& 0.18& 21.60& 0.11& 1.29\ & 1619.29& 420.51& 21.80& 
0.07& 22.09& 0.12& -0.29\nl
   1619.33& 365.93& 22.88& 0.14& 22.57& 0.20& 0.30\ & 1619.33& 501.62& 23.21& 
0.19& 18.64& 0.13& 4.57\nl
   1619.85& 525.60& 20.49& 0.03& 20.92& 0.06& -0.43\ & 1620.04& 358.58& 23.60& 
0.36& 21.93& 0.11& 1.67\nl
   1620.06& 515.55& 20.94& 0.03& 19.46& 0.02& 1.48\ & 1622.35& 657.77& 22.68& 
0.15& 21.93& 0.12& 0.76\nl
   1623.34& 503.97& 22.60& 0.13& 21.07& 0.07& 1.53\ & 1623.70& 538.16& 21.02& 
0.04& 19.96& 0.02& 1.06\nl
   1624.02& 628.90& 21.20& 0.03& 22.05& 0.19& -0.84\ & 1625.04& 400.34& 21.44& 
0.05& 21.87& 0.10& -0.43\nl
   1625.29& 472.84& 21.89& 0.07& 20.55& 0.04& 1.34\ & 1625.76& 311.68& 21.07& 
0.04& 19.86& 0.03& 1.21\nl
   1626.49& 548.62& 22.64& 0.14& 20.38& 0.04& 2.26\ & 1626.74& 426.13& 21.87& 
0.07& 20.38& 0.03& 1.49\nl
   1626.77& 448.50& 21.44& 0.05& 21.25& 0.06& 0.20\ & 1626.82& 622.56& 22.54& 
0.13& 20.72& 0.05& 1.81\nl
   1627.65& 583.32& 18.28& 0.01& 18.29& 0.01& -0.01\ & 1627.79& 490.46& 23.78& 
0.39& 21.99& 0.12& 1.80\nl
   1631.80& 319.54& 17.64& 0.01& 16.84& 0.01& 0.80\ & 1632.15& 459.23& 22.44& 
0.13& 21.34& 0.07& 1.09\nl
   1632.35& 659.54& 23.16& 0.27& 21.70& 0.13& 1.46\ & 1633.76& 598.94& 21.06& 
0.04& 21.33& 0.06& -0.28\nl
   1633.95& 200.21& 23.92& 0.43& 23.12& 0.35& 0.80\ & 1635.44& 502.43& 20.85& 
0.03& 18.90& 0.01& 1.95\nl
   1637.29& 561.73& 20.82& 0.03& 20.83& 0.04& 0.00\ & 1637.64& 548.45& 21.53& 
0.05& 22.84& 0.34& -1.31\nl
   1637.76& 635.98& 22.35& 0.12& 21.43& 0.09& 0.92\ & 1638.37& 450.93& 20.85& 
0.03& 20.74& 0.04& 0.11\nl
   1638.94& 444.23& 22.08& 0.09& 21.76& 0.09& 0.32\ & 1640.83& 570.70& 23.05& 
0.24& 21.35& 0.10& 1.70\nl
   1640.98& 479.22& 21.91& 0.07& 21.77& 0.09& 0.14\ & 1641.22& 521.45& 22.94& 
0.20& 22.54& 0.22& 0.40\nl
   1641.81& 509.58& 22.65& 0.13& 21.49& 0.07& 1.16\ & 1642.19& 610.97& 20.16& 
0.05& 19.12& 0.03& 1.04\nl
   1642.53& 544.93& 21.81& 0.06& 20.79& 0.04& 1.02\ & 1645.73& 663.45& 20.77& 
0.03& 19.44& 0.02& 1.33\nl
   1648.02& 574.60& 23.03& 0.20& 22.26& 0.15& 0.77\ & 1648.89& 629.68& 22.48& 
0.12& 21.16& 0.08& 1.32\nl
   1650.31& 542.19& 22.22& 0.12& 20.42& 0.04& 1.80\ & 1652.04& 505.71& 22.25& 
0.10& 21.02& 0.05& 1.23\nl
   1653.38& 526.92& 22.45& 0.13& 21.45& 0.09& 1.00\ & 1653.55& 628.22& 22.22& 
0.10& 21.00& 0.06& 1.22\nl
   1654.17& 581.97& 21.98& 0.13& 21.23& 0.09& 0.75\ & 1654.24& 385.97& 22.70& 
0.15& 22.32& 0.17& 0.38\nl
   1654.64& 413.47& 23.84& 0.42& 21.87& 0.12& 1.98\ & 1656.38& 541.01& 21.81& 
0.10& 20.81& 0.07& 1.00\nl
   1656.92& 346.66& 22.97& 0.15& 22.56& 0.20& 0.42\ & 1657.44& 649.40& 22.38& 
0.10& 22.22& 0.18& 0.15\nl
   1657.98& 438.44& 22.68& 0.14& 22.53& 0.21& 0.15\ & 1658.05& 554.95& 19.36& 
0.01& 19.59& 0.02& -0.23\nl
   1658.61& 594.12& 22.12& 0.09& 21.90& 0.10& 0.23\ & 1660.90& 662.05& 21.82& 
0.07& 22.32& 0.18& -0.50\nl
   1660.90& 549.56& 20.23& 0.03& 20.17& 0.05& 0.06\ & 1662.66& 647.53& 21.57& 
0.06& 21.64& 0.11& -0.07\nl
   1663.27& 620.04& 22.30& 0.12& 21.67& 0.11& 0.63\ & 1663.66& 589.66& 21.20& 
0.06& 20.12& 0.04& 1.08\nl
   1664.07& 536.82& 22.26& 0.11& 22.43& 0.19& -0.17\ & 1664.83& 493.59& 20.74& 
0.03& 21.16& 0.06& -0.42\nl
   1665.10& 637.64& 20.79& 0.03& 21.25& 0.06& -0.47\ & 1665.50& 545.39& 22.57& 
0.15& 21.50& 0.11& 1.06\nl
   1667.62& 360.90& 22.48& 0.11& 21.68& 0.09& 0.80\ & 1667.83& 465.83& 23.19& 
0.16& 21.90& 0.09& 1.28\nl
   1668.28& 554.67& 20.24& 0.05& 19.88& 0.02& 0.35\ & 1669.36& 444.50& 23.25& 
0.20& 21.50& 0.07& 1.75\nl
   1671.24& 584.83& 22.48& 0.34& 21.63& 0.13& 0.85\ & 1671.31& 463.08& 23.33& 
0.26& 22.18& 0.15& 1.15\nl
   1671.34& 599.72& 22.05& 0.10& 22.52& 0.28& -0.47\ & 1671.49& 605.58& 22.01& 
0.11& 21.84& 0.14& 0.17\nl
   1672.06& 460.73& 23.59& 0.34& 22.72& 0.26& 0.88\ & 1672.07& 454.25& 23.22& 
0.19& 22.37& 0.16& 0.85\nl
   1672.46& 503.02& 23.83& 0.39& 21.81& 0.12& 2.02\ & 1672.58& 479.35& 21.02& 
0.04& 19.72& 0.02& 1.30\nl
   1674.47& 663.75& 21.69& 0.09& 20.59& 0.05& 1.10\ & 1676.80& 594.43& 21.83& 
0.10& 22.10& 0.17& -0.27\nl
   1676.99& 485.72& 20.98& 0.03& 19.63& 0.02& 1.35\ & 1679.03& 416.12& 23.01& 
0.21& 21.53& 0.08& 1.48\nl
   1679.87& 487.75& 20.77& 0.03& 19.43& 0.02& 1.34\ & 1680.75& 315.23& 23.23& 
0.21& 21.71& 0.08& 1.52\nl
   1680.97& 584.76& 22.88& 0.19& 22.69& 0.22& 0.19\ & 1681.97& 515.65& 21.59& 
0.05& 22.82& 0.30& -1.23\nl
   1683.01& 571.77& 21.78& 0.06& 21.66& 0.08& 0.12\ & 1684.15& 435.40& 22.03& 
0.09& 21.23& 0.06& 0.80\nl
   1684.20& 507.46& 20.79& 0.02& 19.61& 0.03& 1.18\ & 1686.47& 602.18& 20.47& 
0.07& 20.83& 0.08& -0.36\nl
   1686.70& 331.71& 22.23& 0.08& 21.19& 0.06& 1.03\ & 1686.78& 491.88& 21.30& 
0.05& 20.08& 0.03& 1.22\nl
   1687.74& 618.46& 21.43& 0.08& 20.47& 0.06& 0.96\ & 1689.05& 630.62& 21.84& 
0.09& 21.52& 0.10& 0.32\nl
   1690.77& 265.77& 23.09& 0.18& 22.51& 0.18& 0.58\ & 1690.83& 567.96& 21.25& 
0.09& 20.84& 0.09& 0.41\nl
   1691.14& 520.84& 20.12& 0.02& 20.54& 0.03& -0.42\ & 1692.42& 600.79& 20.24& 
0.06& 20.45& 0.05& -0.20\nl
   1694.00& 657.57& 21.41& 0.05& 20.24& 0.04& 1.17\ & 1695.82& 481.95& 23.25& 
0.26& 20.79& 0.05& 2.46\nl
   1696.00& 436.79& 21.64& 0.07& 22.42& 0.19& -0.78\ & 1696.43& 442.96& 21.65& 
0.06& 21.51& 0.10& 0.13\nl
   1696.67& 532.62& 20.81& 0.03& 19.02& 0.01& 1.79\ & 1697.42& 551.70& 21.68& 
0.09& 21.33& 0.09& 0.35\nl
   1698.33& 572.92& 22.10& 0.10& 20.79& 0.08& 1.32\ & 1700.68& 507.18& 20.72& 
0.02& 21.38& 0.07& -0.67\nl
   1700.71& 475.52& 23.03& 0.23& 22.47& 0.17& 0.56\ & 1700.91& 521.83& 21.26& 
0.04& 20.03& 0.02& 1.23\nl
   1701.43& 637.32& 22.52& 0.15& 21.67& 0.10& 0.85\ & 1702.78& 664.11& 21.77& 
0.09& 21.02& 0.08& 0.74\nl
   1703.80& 575.51& 21.44& 0.05& 20.17& 0.04& 1.27\ & 1703.82& 556.43& 21.28& 
0.07& 20.94& 0.08& 0.34\nl
   1704.01& 592.68& 21.01& 0.06& 19.83& 0.04& 1.18\ & 1705.09& 617.06& 22.09& 
0.10& 21.08& 0.08& 1.01\nl
   1705.83& 445.17& 23.86& 0.47& 22.21& 0.16& 1.65\ & 1707.13& 491.33& 20.64& 
0.02& 19.45& 0.02& 1.19\nl
   1707.31& 623.29& 21.65& 0.09& 21.08& 0.06& 0.57\ & 1707.32& 552.10& 21.75& 
0.10& 17.90& 0.07& 3.84\nl
   1707.98& 599.31& 19.78& 0.03& 18.12& 0.02& 1.65\ & 1709.04& 642.28& 21.93& 
0.10& 20.99& 0.07& 0.94\nl
   1709.19& 570.80& 21.03& 0.04& 21.26& 0.08& -0.23\ & 1709.22& 586.85& 21.18& 
0.07& 20.09& 0.07& 1.09\nl
   1709.55& 462.81& 22.55& 0.12& 21.45& 0.10& 1.10\ & 1711.52& 620.80& 21.81& 
0.11& 21.48& 0.09& 0.33\nl
   1711.60& 373.71& 22.24& 0.08& 21.44& 0.08& 0.80\ & 1712.71& 553.22& 20.38& 
0.02& 20.58& 0.04& -0.20\nl
   1712.82& 575.47& 21.82& 0.09& 21.41& 0.08& 0.41\ & 1712.87& 636.74& 22.29& 
0.14& 21.30& 0.09& 0.99\nl
   1712.88& 561.48& 22.03& 0.11& 20.51& 0.04& 1.52\ & 1713.10& 605.28& 20.72& 
0.08& 19.90& 0.08& 0.82\nl
   1713.14& 628.97& 21.22& 0.04& 21.38& 0.07& -0.17\ & 1713.29& 172.81& 18.81& 
0.01& 17.50& 0.01& 1.31\nl
   1714.41& 527.65& 21.78& 0.09& 21.48& 0.10& 0.31\ & 1714.97& 363.62& 23.65& 
0.36& 23.36& 0.47& 0.29\nl
   1715.43& 447.52& 23.13& 0.22& 21.95& 0.14& 1.18\ & 1715.46& 326.92& 23.85& 
0.38& 22.80& 0.24& 1.04\nl
   1716.40& 424.20& 21.99& 0.11& 21.50& 0.09& 0.49\ & 1718.97& 539.17& 20.73& 
0.07& 20.65& 0.07& 0.08\nl
   1719.00& 501.54& 22.93& 0.16& 21.16& 0.06& 1.77\ & 1720.57& 653.66& 23.04& 
0.18& 21.84& 0.13& 1.20\nl
   1720.79& 590.41& 20.52& 0.10& 15.87& 0.05& 4.65\ & 1721.85& 551.24& 22.74& 
0.16& 22.33& 0.19& 0.41\nl
   1722.53& 637.60& 22.53& 0.19& 22.16& 0.14& 0.37\ & 1722.96& 380.43& 21.47& 
0.04& 20.89& 0.06& 0.58\nl
   1723.18& 516.24& 20.28& 0.05& 19.05& 0.01& 1.22\ & 1723.20& 526.21& 21.76& 
0.11& 22.13& 0.14& -0.36\nl
   1725.21& 618.67& 22.13& 0.09& 21.64& 0.13& 0.49\ & 1726.25& 628.65& 23.56& 
0.31& 21.69& 0.11& 1.87\nl
   1726.43& 463.43& 22.36& 0.11& 22.02& 0.11& 0.34\ & 1726.56& 651.02& 22.42& 
0.09& 21.33& 0.07& 1.09\nl
   1726.92& 595.16& 20.15& 0.08& 19.09& 0.04& 1.06\ & 1728.26& 486.56& 21.97& 
0.08& 21.06& 0.07& 0.92\nl
   1728.86& 530.03& 21.34& 0.04& 21.09& 0.07& 0.25\ & 1729.28& 624.73& 21.71& 
0.08& 21.11& 0.08& 0.60\nl
   1729.33& 615.37& 22.26& 0.13& 21.01& 0.09& 1.25\ & 1729.35& 604.91& 20.23& 
0.06& 19.40& 0.03& 0.84\nl
   1732.03& 648.37& 23.02& 0.17& 21.65& 0.12& 1.37\ & 1733.74& 621.98& 21.13& 
0.05& 20.98& 0.07& 0.15\nl
   1734.40& 488.56& 21.45& 0.04& 21.05& 0.07& 0.40\ & 1735.63& 565.53& 21.56& 
0.05& 21.91& 0.13& -0.35\nl
   1736.58& 593.56& 20.34& 0.06& 19.50& 0.04& 0.84\ & 1738.37& 343.45& 23.11& 
0.17& 22.26& 0.14& 0.85\nl
   1738.92& 270.54& 23.49& 0.30& 22.19& 0.16& 1.30\ & 1739.06& 643.63& 22.16& 
0.13& 21.80& 0.14& 0.35\nl
   1740.36& 422.98& 23.09& 0.22& 22.93& 0.28& 0.16\ & 1740.64& 242.81& 20.86& 
0.03& 18.73& 0.02& 2.13\nl
   1741.08& 639.64& 22.09& 0.07& 21.53& 0.11& 0.57\ & 1741.81& 259.60& 23.67& 
0.41& 23.56& 0.46& 0.11\nl
   1742.73& 533.22& 23.40& 0.30& 23.57& 0.54& -0.18\ & 1745.49& 511.50& 23.42& 
0.28& 22.75& 0.27& 0.67\nl
   1746.27& 325.36& 22.42& 0.11& 22.40& 0.19& 0.02\ & 1746.56& 414.51& 23.28& 
0.21& 21.39& 0.15& 1.89\nl
\enddata
\end {planotable}

\begin{planotable}{rrrrrrrrrrrrrr}
\small
\footnotesize
\tablewidth{0pt}
\scriptsize
\tablecaption{Photometry of Background Field}
\tablehead{
\multicolumn{1}{c}{X}&
\multicolumn{1}{c}{Y}&
\multicolumn{1}{c}{V}&
\multicolumn{1}{c}{$\sigma_V$}&
\multicolumn{1}{c}{I}&
\multicolumn{1}{c}{$\sigma_I$}&
\multicolumn{1}{c}{$V-I$}&
\multicolumn{1}{c}{X}&
\multicolumn{1}{c}{Y}&
\multicolumn{1}{c}{V}&
\multicolumn{1}{c}{$\sigma_V$}&
\multicolumn{1}{c}{I}&
\multicolumn{1}{c}{$\sigma_I$}&
\multicolumn{1}{c}{$V-I$}}
\startdata
    246.81&  33.30& 16.84& 0.01& 16.35& 0.01& 0.49\ & 1634.89& 122.69& 16.59& 
0.02& 15.91& 0.02& 0.67\nl
   1555.28&  60.56& 16.42& 0.03& 15.61& 0.05& 0.81\ & 1670.17&  55.69& 16.08& 
0.04& 15.52& 0.05& 0.56\nl
    391.71& 116.74& 16.34& 0.05& 15.64& 0.04& 0.70\ & 92.11& 307.73& 15.39& 
0.07& 14.91& 0.05& 0.49\nl
    431.19& 387.44& 16.19& 0.06& 15.79& 0.04& 0.41\ & 5.21& 101.15& 21.71& 
0.05& 20.31& 0.04& 1.40\nl
      9.06& 490.30& 23.17& 0.19& 21.41& 0.08& 1.76\ & 13.20& 393.45& 18.83& 
0.01& 18.16& 0.01& 0.67\nl
     25.90& 495.50& 24.28& 0.50& 23.16& 0.39& 1.12\ & 27.46&  27.37& 22.04& 
0.12& 21.46& 0.10& 0.59\nl
     41.99& 210.82& 22.24& 0.09& 21.62& 0.09& 0.62\ & 42.86& 154.97& 20.94& 
0.02& 19.55& 0.01& 1.39\nl
     47.54& 310.15& 22.77& 0.12& 21.79& 0.11& 0.98\ & 53.11& 477.84& 23.69& 
0.35& 22.27& 0.16& 1.41\nl
     58.57& 370.44& 24.03& 0.44& 23.01& 0.39& 1.02\ & 60.23& 383.35& 23.08& 
0.20& 21.84& 0.12& 1.24\nl
     64.09& 575.36& 21.27& 0.07& 19.89& 0.05& 1.38\ & 79.21& 375.53& 21.25& 
0.03& 20.83& 0.05& 0.42\nl
     81.75&   1.47& 22.41& 0.14& 21.17& 0.10& 1.25\ & 112.50& 452.45& 23.89& 
0.40& 23.17& 0.30& 0.72\nl
    115.43&  86.91& 23.19& 0.21& 22.01& 0.14& 1.18\ & 123.01& 563.98& 19.27& 
0.01& 18.74& 0.01& 0.53\nl
    129.26& 317.15& 20.21& 0.02& 19.74& 0.01& 0.46\ & 130.66& 657.24& 23.63& 
0.26& 22.07& 0.17& 1.56\nl
    141.73& 285.03& 23.67& 0.27& 23.08& 0.34& 0.59\ & 149.70& 451.57& 24.14& 
0.40& 23.19& 0.54& 0.96\nl
    150.38& 277.60& 24.10& 0.43& 21.96& 0.12& 2.14\ & 151.78& 334.16& 23.98& 
0.47& 23.43& 0.51& 0.55\nl
    159.64& 202.24& 18.23& 0.01& 17.56& 0.01& 0.67\ & 161.98& 261.17& 20.41& 
0.02& 19.88& 0.02& 0.53\nl
    175.69& 123.52& 23.86& 0.36& 23.02& 0.30& 0.84\ & 178.66& 518.49& 24.04& 
0.33& 22.97& 0.42& 1.07\nl
    185.41& 471.75& 21.72& 0.05& 20.33& 0.03& 1.38\ & 191.35& 200.32& 24.19& 
0.43& 22.85& 0.39& 1.34\nl
    215.62&  47.76& 24.22& 0.39& 23.32& 0.52& 0.90\ & 222.46& 161.78& 23.63& 
0.46& 23.16& 0.41& 0.47\nl
    228.24& 360.29& 22.93& 0.16& 21.68& 0.10& 1.25\ & 232.59&  96.41& 24.27& 
0.53& 23.64& 0.54& 0.63\nl
    270.70& 468.93& 23.35& 0.19& 21.45& 0.09& 1.90\ & 271.41& 300.32& 23.60& 
0.34& 23.18& 0.44& 0.43\nl
    280.53& 540.91& 23.88& 0.42& 21.68& 0.09& 2.20\ & 289.59& 389.99& 19.95& 
0.01& 18.98& 0.01& 0.97\nl
    290.50& 510.00& 24.19& 0.42& 23.22& 0.45& 0.97\ & 290.92&  82.19& 21.76& 
0.06& 19.22& 0.02& 2.54\nl
    304.43& 405.52& 21.48& 0.07& 19.91& 0.04& 1.57\ & 309.07& 607.91& 24.26& 
0.43& 23.52& 0.50& 0.74\nl
    355.93&  79.71& 24.13& 0.50& 23.27& 0.43& 0.86\ & 356.78& 384.78& 23.05& 
0.19& 21.01& 0.04& 2.04\nl
    360.77& 122.23& 23.49& 0.25& 22.10& 0.12& 1.39\ & 390.10& 538.87& 22.42& 
0.12& 20.67& 0.05& 1.75\nl
    391.85& 573.53& 23.09& 0.18& 21.73& 0.10& 1.36\ & 396.99& 409.93& 17.29& 
0.01& 16.83& 0.01& 0.46\nl
    401.62&  41.31& 17.78& 0.01& 17.21& 0.01& 0.57\ & 417.64& 165.77& 23.82& 
0.44& 22.84& 0.31& 0.99\nl
    417.67& 264.74& 22.59& 0.12& 20.81& 0.05& 1.78\ & 446.92& 227.94& 20.93& 
0.03& 20.51& 0.03& 0.42\nl
    453.29& 423.30& 23.72& 0.28& 21.95& 0.11& 1.77\ & 480.55& 545.63& 21.55& 
0.05& 19.36& 0.02& 2.19\nl
    487.36& 413.35& 18.08& 0.01& 16.74& 0.01& 1.35\ & 501.17& 269.66& 23.21& 
0.20& 21.13& 0.06& 2.08\nl
    507.17& 291.03& 22.45& 0.09& 20.59& 0.04& 1.86\ & 509.84& 522.19& 20.44& 
0.02& 20.04& 0.02& 0.40\nl
    512.25& 266.67& 22.59& 0.12& 21.10& 0.06& 1.49\ & 574.02& 477.68& 23.66& 
0.28& 22.57& 0.20& 1.08\nl
    593.60& 549.77& 23.49& 0.24& 23.03& 0.36& 0.46\ & 600.03& 344.42& 19.52& 
0.01& 18.72& 0.01& 0.80\nl
    610.64& 248.81& 24.27& 0.53& 21.52& 0.08& 2.75\ & 618.29& 244.10& 20.54& 
0.02& 19.80& 0.02& 0.75\nl
    620.63& 339.48& 22.93& 0.12& 20.34& 0.03& 2.60\ & 623.14& 650.07& 22.85& 
0.13& 20.94& 0.04& 1.91\nl
    637.97& 613.74& 21.11& 0.08& 19.63& 0.06& 1.48\ & 640.10& 302.71& 20.54& 
0.02& 18.65& 0.01& 1.89\nl
    640.19& 646.47& 23.26& 0.22& 22.47& 0.14& 0.79\ & 642.71& 228.96& 24.16& 
0.49& 22.47& 0.22& 1.69\nl
    665.26& 151.82& 22.03& 0.09& 20.84& 0.05& 1.19\ & 666.05&  40.78& 23.03& 
0.21& 22.52& 0.20& 0.51\nl
    666.67& 155.68& 21.80& 0.08& 20.81& 0.05& 0.99\ & 667.02& 535.41& 21.35& 
0.04& 18.84& 0.01& 2.50\nl
    669.70& 287.57& 18.66& 0.01& 18.17& 0.01& 0.49\ & 673.54& 385.42& 23.80& 
0.30& 22.18& 0.14& 1.61\nl
    677.74& 401.53& 21.47& 0.04& 20.12& 0.02& 1.35\ & 699.32&  29.97& 21.79& 
0.12& 20.51& 0.07& 1.28\nl
    701.78& 368.59& 21.58& 0.09& 20.27& 0.07& 1.30\ & 703.15& 629.78& 24.18& 
0.40& 23.28& 0.41& 0.90\nl
    747.47& 412.66& 24.17& 0.46& 23.52& 0.52& 0.65\ & 751.14& 158.61& 23.81& 
0.37& 22.27& 0.16& 1.54\nl
    754.47& 214.54& 23.80& 0.37& 22.73& 0.28& 1.07\ & 755.32&   2.75& 22.88& 
0.21& 21.59& 0.08& 1.29\nl
    764.12& 134.73& 24.11& 0.37& 22.90& 0.44& 1.21\ & 766.08& 249.02& 22.65& 
0.12& 21.77& 0.09& 0.88\nl
    776.65& 437.04& 22.97& 0.17& 20.66& 0.04& 2.31\ & 810.07& 122.05& 24.14& 
0.48& 23.46& 0.41& 0.68\nl
    817.75& 477.48& 23.55& 0.25& 21.95& 0.10& 1.59\ & 843.14&  40.73& 21.53& 
0.05& 20.55& 0.04& 0.98\nl
    846.03& 310.12& 22.53& 0.11& 21.44& 0.06& 1.09\ & 851.23& 543.52& 22.88& 
0.19& 21.87& 0.11& 1.01\nl
    866.87& 545.86& 21.80& 0.09& 21.17& 0.09& 0.63\ & 878.27& 532.91& 23.08& 
0.19& 20.98& 0.05& 2.10\nl
    889.86& 570.06& 24.00& 0.40& 23.34& 0.41& 0.66\ & 896.98& 618.05& 22.51& 
0.12& 22.50& 0.18& 0.01\nl
    901.38& 171.38& 22.14& 0.07& 20.89& 0.05& 1.25\ & 908.05& 364.26& 20.43& 
0.02& 18.83& 0.01& 1.59\nl
    917.76&  57.97& 17.40& 0.01& 16.31& 0.01& 1.09\ & 918.08& 469.86& 19.72& 
0.01& 18.96& 0.01& 0.76\nl
    940.85& 419.43& 23.25& 0.19& 22.14& 0.16& 1.10\ & 955.46& 656.59& 23.75& 
0.31& 21.83& 0.13& 1.92\nl
    956.05& 320.87& 23.92& 0.40& 23.39& 0.46& 0.53\ & 956.79& 654.89& 23.77& 
0.37& 21.82& 0.12& 1.95\nl
    958.15& 380.78& 19.38& 0.01& 18.84& 0.01& 0.54\ & 968.29& 207.51& 20.54& 
0.02& 20.10& 0.02& 0.44\nl
    972.06& 181.21& 22.67& 0.10& 21.43& 0.06& 1.24\ & 974.82& 405.69& 23.70& 
0.32& 22.05& 0.11& 1.65\nl
    976.69& 486.20& 23.70& 0.33& 22.84& 0.25& 0.86\ & 989.01&  90.57& 23.79& 
0.37& 21.18& 0.05& 2.61\nl
   1005.43&  68.97& 21.83& 0.09& 20.46& 0.05& 1.37\ & 1005.67&  23.52& 21.88& 
0.11& 20.22& 0.06& 1.66\nl
   1010.28&  61.51& 21.85& 0.06& 19.81& 0.03& 2.04\ & 1013.87& 522.05& 24.19& 
0.48& 23.51& 0.51& 0.68\nl
   1023.80&  70.93& 23.65& 0.25& 22.16& 0.19& 1.49\ & 1023.90& 518.10& 23.39& 
0.25& 21.37& 0.06& 2.01\nl
   1029.53& 619.88& 21.56& 0.04& 21.15& 0.06& 0.41\ & 1030.44& 475.40& 23.74& 
0.37& 22.31& 0.19& 1.43\nl
   1035.16& 416.25& 22.58& 0.12& 21.60& 0.07& 0.97\ & 1039.53& 610.96& 22.06& 
0.08& 19.49& 0.01& 2.57\nl
   1045.45& 448.48& 23.85& 0.34& 22.60& 0.17& 1.25\ & 1057.47& 588.77& 23.90& 
0.33& 22.98& 0.32& 0.92\nl
   1076.08& 344.36& 23.82& 0.37& 23.42& 0.49& 0.41\ & 1079.56& 601.90& 23.27& 
0.21& 21.43& 0.08& 1.84\nl
   1087.40& 438.94& 22.61& 0.13& 21.87& 0.09& 0.75\ & 1099.23& 261.87& 23.60& 
0.29& 23.07& 0.32& 0.53\nl
   1099.47& 102.35& 23.86& 0.30& 21.87& 0.10& 2.00\ & 1117.27&  95.55& 18.38& 
0.00& 17.86& 0.01& 0.51\nl
   1128.81& 215.10& 23.85& 0.38& 22.01& 0.11& 1.84\ & 1134.04&   8.74& 23.98& 
0.47& 21.61& 0.08& 2.36\nl
   1157.39&  86.25& 23.98& 0.35& 22.45& 0.21& 1.53\ & 1167.58&  19.61& 19.77& 
0.01& 18.26& 0.01& 1.51\nl
   1185.34&  37.68& 22.85& 0.16& 21.28& 0.07& 1.57\ & 1292.42&  24.57& 24.05& 
0.50& 21.79& 0.10& 2.26\nl
   1292.57&  22.77& 23.92& 0.44& 21.80& 0.10& 2.12\ & 1537.64&  68.94& 23.73& 
0.36& 22.79& 0.22& 0.94\nl
   1608.40&  66.13& 24.08& 0.49& 23.51& 0.49& 0.56\ & 1626.19&   5.55& 22.29& 
0.14& 21.64& 0.09& 0.65\nl
\enddata
\end {planotable}

\end{document}